\documentclass[preprint,aps]{revtex4-1}

\usepackage{graphicx}
\usepackage{dcolumn}
\usepackage{bm}
\usepackage{amsmath,amssymb}
\usepackage{subfigure}
\usepackage{color}
\usepackage{hyperref}
\usepackage{rotating}
\usepackage[normalem]{ulem}
\usepackage{gensymb}

\usepackage [english]{babel}
\usepackage [autostyle, english = american]{csquotes}
\usepackage{array}

\definecolor{maroon}{RGB}{102,0,204}
\definecolor{darkgreen}{RGB}{0,102,51}
\definecolor{orange}{RGB}{255,128,0}
\definecolor{brown}{RGB}{153,76,0}
\definecolor{light-gray}{gray}{0.5}
\definecolor{red}{rgb}{1.0, 0.0, 0.0}

\begin{document}


\title{
   Hydrodynamics of Flapping Foils Undergoing Irregular Motion with Application to  Wave-Assisted Propulsion
 }
\author{Harshal S. Raut}
 \email{hraut1@jhu.edu}
\affiliation{Department of Mechanical Engineering, Johns Hopkins University, Baltimore, MD, USA}%
\author{Jung-Hee Seo}
 \email{jhseo@jhu.edu}
 \affiliation{Department of Mechanical Engineering, Johns Hopkins University, Baltimore, MD, USA}%
\author{Rajat Mittal}%
 \email{mittal@jhu.edu}
\affiliation{Department of Mechanical Engineering, Johns Hopkins University, Baltimore, MD, USA}%

\date{\today}

\begin{abstract}
Flapping foils are widely studied as bioinspired propulsors, yet most investigations have focused on regular, sinusoidal kinematics. In realistic environments, however, irregular motions arise naturally due to environmental disturbances, fluid-structure interactions, and control inputs, but their hydrodynamic consequences remain largely unexplored. One system where response to irregular forcing is particularly relevant is wave-assisted propulsion (WAP) systems where free-to-pitch submerged foils generate thrust due to wave-induced heaving. We employ time-accurate flow simulations of elliptic WAP foils subjected to irregular waves at three different sea-states to gain insights into this system. Our results demonstrate that irregular heaving and pitching can generate greater mean thrust than energetically equivalent sinusoidal heaving. Moreover, a spring-based pitch-limiting mechanism yields higher thrust than an angle-limiter under the same conditions. By leveraging a previously developed leading-edge vortex (LEV) model, we uncover the mechanisms driving this thrust enhancement and highlight the critical interactions between unsteady flow structures and foil dynamics. These findings provide new insights into flapping-foil propulsion in irregular environments and have direct implications for the design and optimization of WAP systems. 

\end{abstract}

\maketitle
\section{Introduction} \label{sec:intro}
Although flapping foils have been extensively investigated under sinusoidal or otherwise periodic motions \cite{TRIANTAFYLLOU1993205, anderson1998oscillating, wu2020review}, relatively little attention has been devoted to irregular or aperiodic flapping. This gap is notable because many biological propulsors, including fish, birds, and insects, rarely maintain strictly periodic kinematics in natural locomotion. Instead, their flapping motions are often perturbed by environmental disturbances, unsteady body dynamics, or active control strategies, resulting in inherently irregular (non-periodic) patterns \cite{srygley2002unconventional, lauder2002forces, dickinson1999wing}. The flow physics associated with such irregular motion are expected to differ significantly from sinusoidal motions, with short bursts of large acceleration and rapid changes in effective angle-of-attack likely dominating thrust and lift generation. Understanding these unsteady mechanisms is essential for linking biological observations to flow physics and for developing robust bioinspired propulsion concepts that can operate efficiently and even harness the energy in realistic environments.

\begin{figure}
    \includegraphics[width=0.8\textwidth]{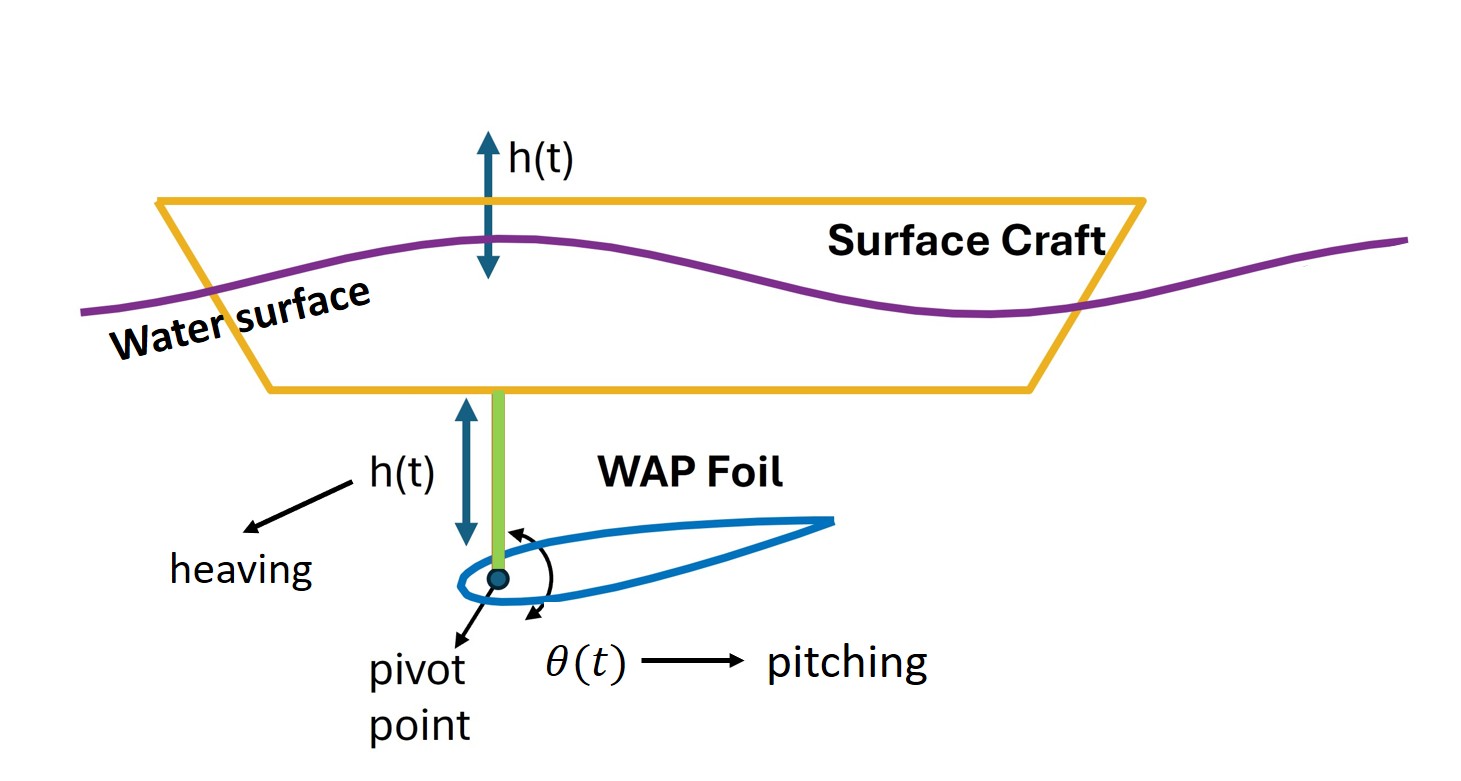}
    \caption{Schematic of surface craft with a wave-assisted propulsion (WAP) system. The schematic is not drawn to scale.}
    \label{WAP_Vessel}
\end{figure}
One system where the performance of flapping foils undergoing irregular motion is particularly relevant is for wave-assisted propulsion (WAP) systems which harness the energy of ocean waves via flapping foils to generate thrust for surface vessels. A range of WAP configurations have been proposed and examined \cite{xing2023wave,zhang2024experimental,zhang2022wave,xu2024analysis,qi2020effect,zhang2024dual,yang2019systematic}, most of which employ a hydrofoil mounted beneath the hull and connected to the supporting strut through a torsional spring (Fig. \ref{WAP_Vessel}). Vertical oscillations of the vessel induced by surface waves drive a periodic heaving motion of the submerged foil. The coupling between hydrodynamic forces and the restoring moment of the torsional spring results in a passive pitching response. The combined heaving and pitching kinematics generate thrust, which can function either as the main propulsion mechanism or as a supplementary contribution to the vessel’s conventional propulsion system.

\begin{figure}
    \centering
    \subfigure[]
    {\includegraphics[width=0.44\textwidth,trim={0cm 0cm 0cm 0cm},clip]{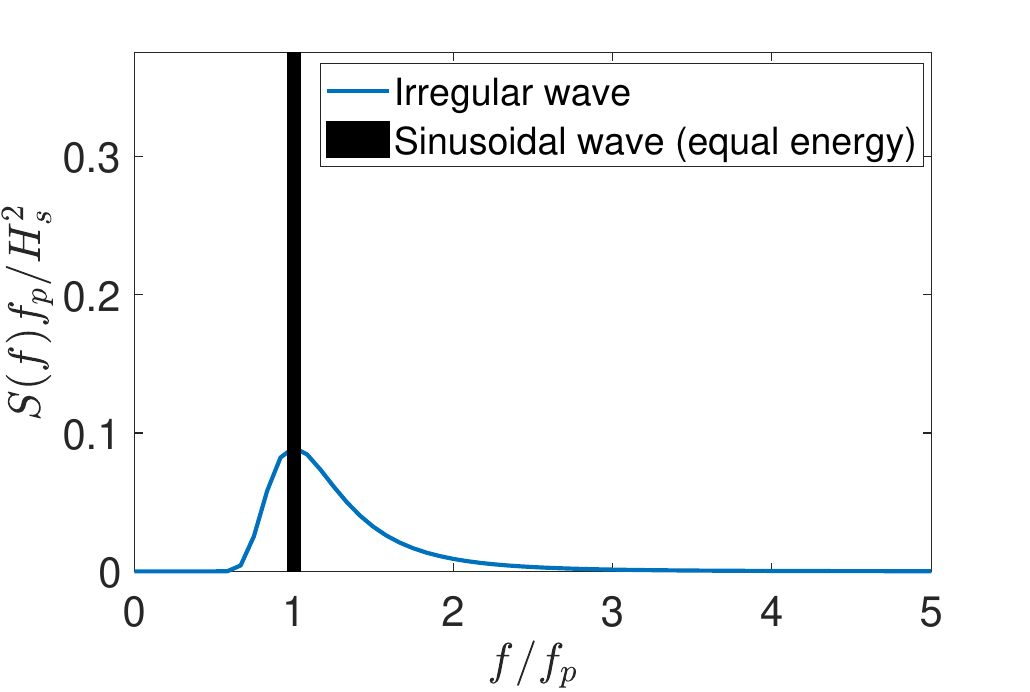}}
    \subfigure[]
    {\includegraphics[width=0.53\textwidth,trim={0cm 0cm 0cm 0cm},clip]{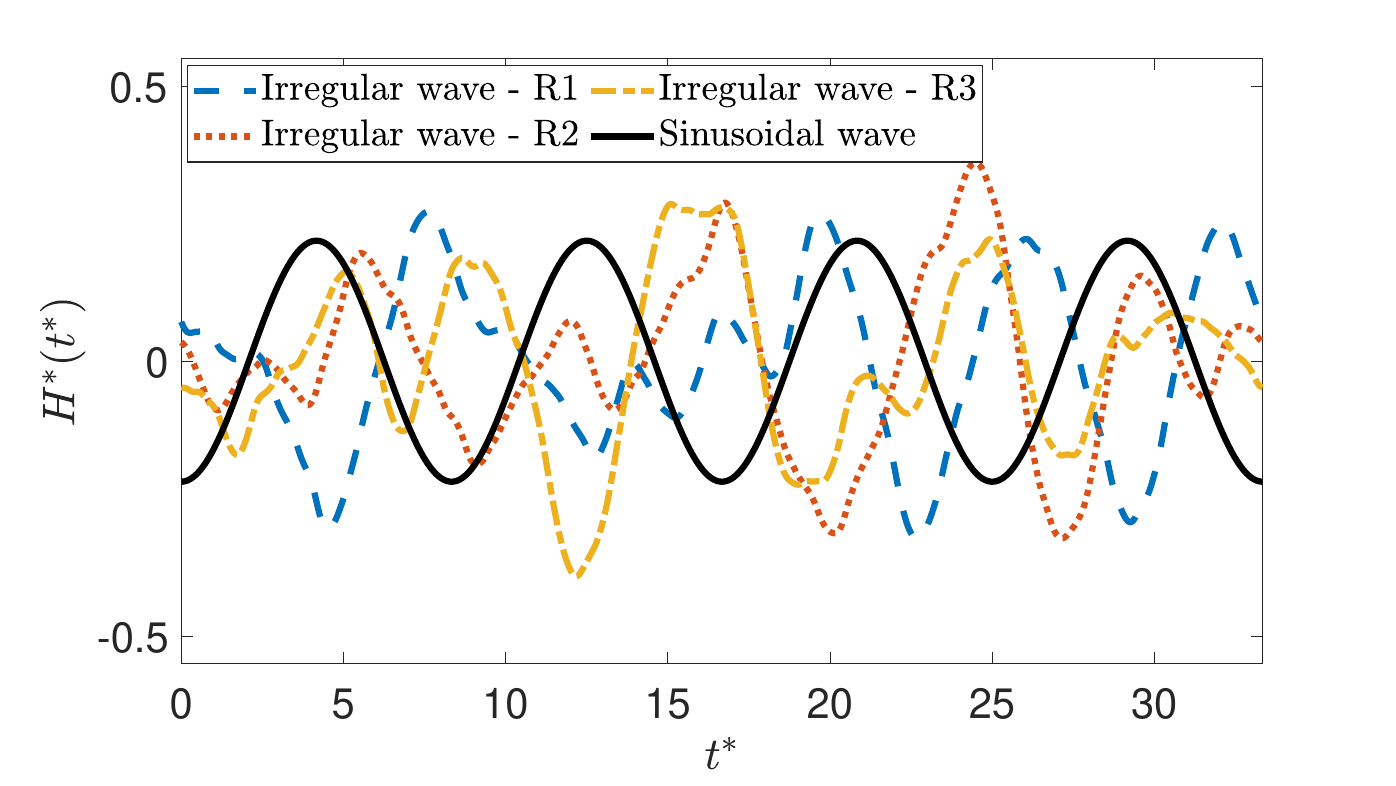}}
    \caption{(a) Non-dimensional Bretschneider wave energy spectrum for irregular waves. (b) Three realizations (R1, R2 and R3) of irregular waves generated from the Bretschneider specturm for $H_s^*=0.62$ and St$_C=0.12$.}
    \label{Bret_intro}
\end{figure}

In realistic ocean environments, wave fields are inherently irregular, exhibiting variability in frequency, amplitude, wavelength, and direction. Such irregularity is typically represented through statistical descriptions in the form of wave spectra, which specify the distribution of wave energy across frequencies. Among the commonly adopted spectral models, the Bretschneider spectrum is widely used to characterize fully developed seas \cite{bretschneider1959wave}, whereas the JONSWAP spectrum provides a more accurate representation of developing seas, particularly in coastal and limited-fetch regions \cite{hasselmann1973measurements}. In this study, irregular waves are modeled using the Bretschneider spectrum. Fig. \ref{Bret_intro}(a) shows the non-dimensional form of the Bretschneider spectrum along with a reference regular wave included for comparison. The reference regular wave is defined such that its total energy equals the integrated energy of the irregular spectrum,
\begin{equation}
    E_\text{irr} = \rho g \int_0^{\infty} S(f)df = \rho g \frac{H_s^2}{16}
    \label{E_irregular}
\end{equation}
where $S(f)$ is the spectral energy density and $H_s$ represents significant wave height which is a standardized statistical value representing the average height of the highest one-third of waves in a sea-state. For a monochromatic regular (i.e. sinusoidal) wave of amplitude $A$, the average wave energy per unit surface area is given by
\begin{equation}
    E_\text{reg} = \frac{1}{8} \rho g A^2.
    \label{E_regular}
\end{equation}
Equating $E_\text{irr} = E_\text{reg}$ yields the amplitude $A$ of the equivalent regular wave. To represent this wave consistently within the spectral framework, its energy is concentrated into a single frequency bin of width $\Delta f$ centered at the chosen frequency. Accordingly, the spectral ordinate of the regular wave is determined from the condition
\begin{equation}
    S_\text{reg}(f)\Delta f = E_\text{reg},
\end{equation}
which ensures that the spectral contribution of the regular wave reflects the same total energy as the irregular spectrum. Fig. \ref{Bret_intro}(b) illustrates the stochastic variability of irregular seas by showing three independent realizations of the irregular wave field, plotted together with the equal-energy regular wave for reference.

A few studies have addressed the influence of irregular waves on WAP systems. Filippas and Belibassakis \cite{filippas2014hydrodynamic} analyzed the hydrodynamic performance of actively controlled hydrofoils beneath the free surface and demonstrated their capability to generate thrust under irregular sea conditions represented by the Bretschneider spectrum. In a related study, Politis and Politis \cite{politis2014biomimetic} investigated hydrofoils subjected to irregular heave motions, also modeled with the Bretschneider spectrum, in combination with active pitching. They further proposed a control strategy that dynamically adjusts the pitch angle to improve propulsive efficiency. More recently, Feng \emph{et al.} \cite{feng2025dynamic} developed a reduced-order dynamical model of a wave glider operating in irregular seas described by the JONSWAP spectrum, showing that stiffer torsional springs are required in higher-energy wave environments to maximize thrust. These findings highlight that the variability inherent in irregular seas can cause significant changes in both performance and dynamic response of wave gliders across operational conditions. Despite progress in understanding multi-body motion interactions and spring-stiffness effects, the majority of design approaches still assume regular, well-characterized wave conditions. As a result, the optimal selection of WAP foil parameters for robust performance in realistic, irregular sea-states remains an open research question.

In our earlier work \cite{raut2024hydrodynamic}, we examined the influence of pivot location and torsional spring stiffness on the thrust generation of WAP systems through the use of energy maps, and proposed a predictive model that estimates thrust from foil kinematics based on the phenomenology of leading-edge vortex (LEV) dynamics. Building upon this, we further investigated multi-foil arrangements \cite{raut2025dynamics}, demonstrating how the wake shed by an upstream foil can be exploited to enhance the performance of downstream foils. 

More recently, based on the model prediction that pitch angles that exceed certain magnitudes based on the heave Strouhal number diminish thrust performance, we analyzed two passive pitch-constraining mechanisms for WAP systems —termed the ``spring-limiter'' and the ``angle-limiter''—and assessed their impact on propulsion efficiency in regular sinusoidal waves \cite{raut2025harnessing}. The spring-limiter regulates the pitching motion via a torsional spring, while the angle-limiter simply restricts the maximum pitch amplitude via a mechanical stopper. In addition, we evaluated the influence of foil geometry and identified the thin elliptical foil as the most effective foil geometry. Motivated by these findings, the present study adopts a thin elliptical foil to examine the performance and hydrodynamic characteristics of a WAP foil operating in irregular sea-states.

Given the inherently irregular nature of ocean waves, we employ numerical simulations to investigate their impact on the thrust generation of WAP propulsors. As a baseline, the thrust performance of the system is first examined under both regular and irregular waves of equal energy across a range of sea-states. Subsequently, two passive pitch-constraint mechanisms—the torsional spring-limiter and the angle-limiter—are evaluated to assess their influence on propulsive efficiency. High-fidelity CFD simulations are carried out to quantify the effects of torsional spring stiffness and the maximum pitch angle permitted by the angle-limiter across sea-states characterized by varying wave heights and frequencies. The force partitioning method (FPM) is applied to quantify the contribution of different components of the pressure-induced forces on thrust generation. In total, 128 simulations are performed (32 corresponding to regular waves and 96 to irregular waves, and these are equally divided between the spring-limiter and angle-limiter cases).

These simulations provide a systematic and quantitative assessment of the role of irregular wave forcing and pitch-constraint mechanisms in shaping the thrust performance of WAP systems. The simulations also allow us to explore the flow-physics of foils undergoing irregular flapping motion and in particular, how fluid-structure interaction and system compliance can be harnessed to improve the performance of flapping foil systems in these situations.

\section{\label{setup}Methodology}
\subsection{\label{prob_setup}Problem Set-up}
In the the current study, we employ an elliptic hydrofoil with a 25:2 aspect-ratio. This foil is chosen based on our previous study which showed that the elliptic foil gives best performance for WAP applications \cite{raut2025harnessing}.

\begin{figure}
    \centering
    \subfigure[]
    {\includegraphics[width=0.4\textwidth,trim={0cm 0cm 0cm 0cm},clip]{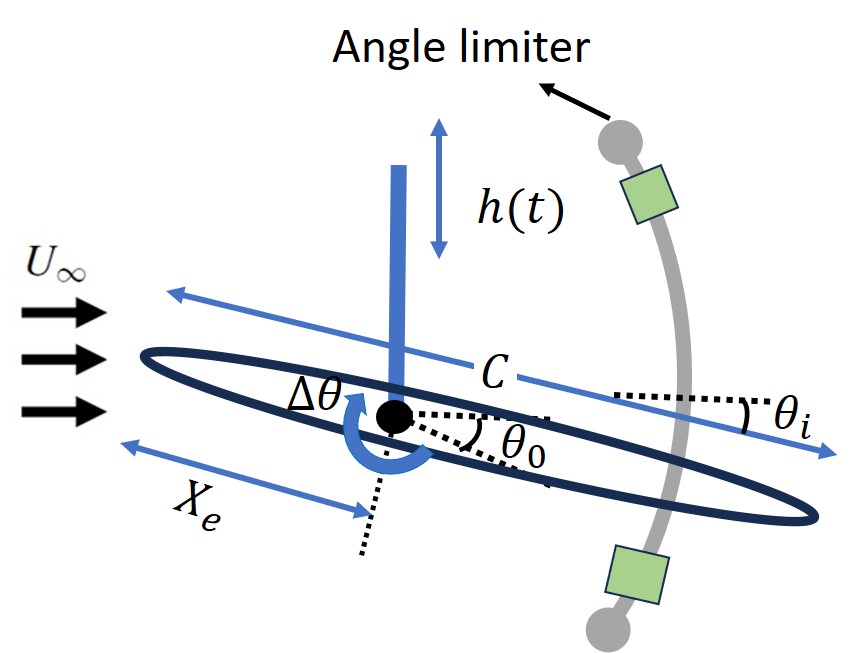}}
    \subfigure[]
    {\includegraphics[width=0.4\textwidth,trim={0cm 0cm 0cm 0cm},clip]{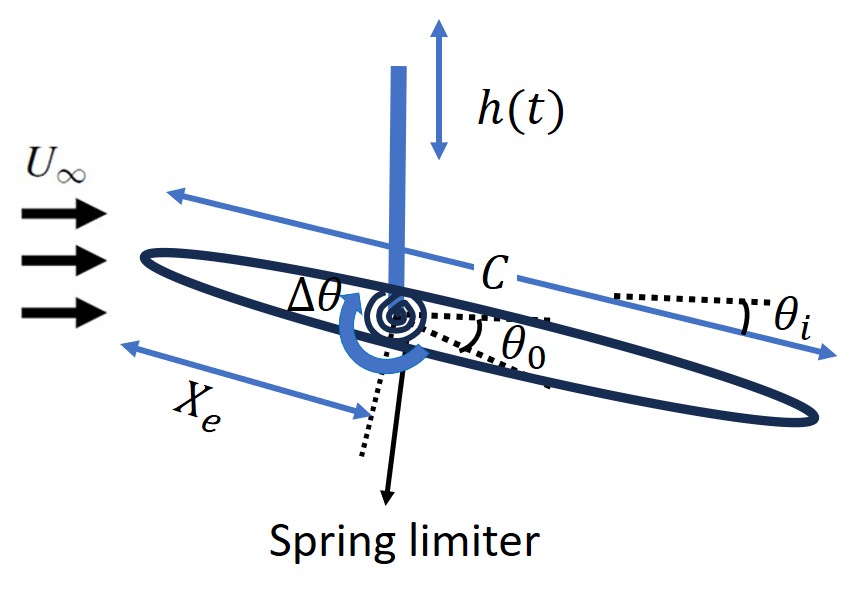}}
    \subfigure[]
    {\includegraphics[width=0.8\textwidth,trim={0cm 0cm 0cm 0cm},clip]{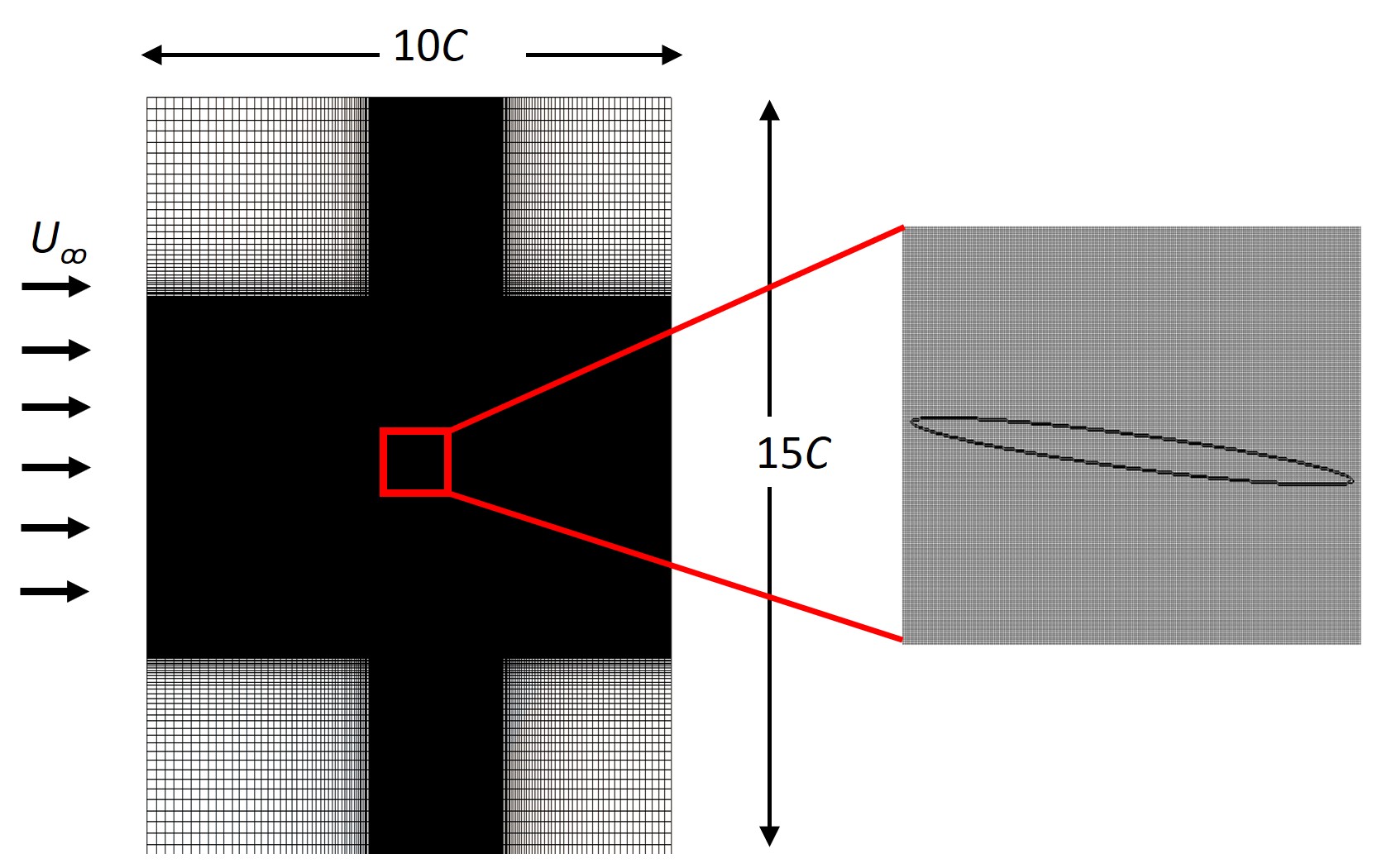}\label{computational_domain}}
    \caption{Schematic of the hydroelastic system used in this study for (a) spring-limiter and (b) angle-limiter. (c) Computational domain and close-up of the Cartesian computational grid.}
    \label{schematic}
\end{figure}

The governing incompressible Navier-Stokes equations expressed in the dimensionless form are as follows:
\begin{align}
    \frac{\partial \textbf{u}}{\partial t} + \textbf{u}.\nabla\textbf{u} &= -\mathbf{\nabla p} + \frac{1}{Re} \nabla^2 \textbf{u} \\
    \nabla . \textbf{u} &= 0
\end{align}
where $Re=\rho U_{\infty}C/\mu$ represents the chord-based Reynolds number. {\color{black}We assume that the hydrofoil is completely submerged and far from the free surface. We therefore do not consider the effect of free surface on the foil}. {\color{black}Thus, the only effect of the surface wave is to provide the heaving motion which is connected with the given sea-state}. In the case of a foil with a torsional spring, pitching is induced through the balance between flow-induced forces and restoring torque by the torsional spring, mounted at the hinge location, illustrated in Fig. \ref{schematic} (a). In the case of the angle-limiter, the foil is allowed to pitch freely due to the flow-induced forces but a constraint is imposed on the maximum allowable pitch angle. This may be implemented simply by a stopper as illustrated in Fig. \ref{schematic} (b).
The linear torsional spring has a spring constant $k_\theta$, and the rotational hinge, or the pivot axis, is at a non-dimensional chordwise location of $X_e^{*} = X_e/C$. The maximum angle allowed by the angle-limiter is denoted as $\theta_{0}$. The equilibrium angular position of the spring is denoted by angle $\theta_i$ which has been kept zero throughout all the simulations in this study. In the case of a spring-limiter, the flow-induced pitching motion is governed by the equation of a forced angular spring-mass oscillator. The equation is non-dimensionalized by the characteristic variables used for the flow (length: $C$; time: $C/U_{\infty}$), and the resulting non-dimensionalized equation governing the hydrofoil's pitch dynamics is given by:
\begin{equation}
    I^* \ddot{\theta}+k^*\theta  = C_M
    \label{solid_equation}
\end{equation}
where $C_M = M/(0.5 \rho U^2_{\infty} C^2 b)$ is the coefficient of pitching moment, and
$I^*=2I/(\rho C^4 b)$, $k^*= 2k/(\rho U^2_{\infty} C^2 b)$ are the dimensionless moment-of-inertia and spring stiffness, respectively. Here, $b$ is the spanwise width of the foil. Both $C_M$ and $I^*$ are calculated with respect to $X_e^*$. The moment of inertia $I$ is calculated using the following volume integral $\rho\int_V r^2 dV$ where $r$ is the radial distance from the pivot location $X_e$. In this work, the torsional spring stiffness is expressed in terms of a non-dimensional frequency ratio $f_{\theta}/f_p$ where $f_{\theta}=(1/2\pi)\sqrt{k/I}$ is the natural frequency of the spring-mass system and $f_p$ is the peak frequency of the heaving motion in the $y$-direction. The above equation can be written in terms of $f_{\theta}$ as
\begin{equation}
    \ddot{\theta}+\Bigg(2\pi \frac{f_\theta}{f_p} \textrm{St}_C\Bigg)^2\theta  = \frac{C_M}{I^*}
    \label{solid_eq2}
\end{equation}
where $\text{St}_C=f_pC/U_\infty$ is the Strouhal number based on chord length. In the case of the angle-limiter, the second term in Eq. \ref{solid_eq2} is omitted and the pitch velocity is set to zero if the maximum allowable angle $\theta_{0}$ is reached and the pitching moment $C_M$ is tending to push the foil beyond the maximum angle. The foil is again allowed to pitch as soon as $C_M$ changes in direction. {\color{black}Structural damping could be present in engineered system, but in order to keep the WAP system simple we have not included it in this study.} 

The prescribed heaving motion in the $y$-direction for the foil, $h(t)$, is modeled as an irregular wave motion generated from the Bretschneider spectrum, a parametric wave spectrum commonly used to represent fully developed sea-states under wind-driven conditions. In non-dimensional form, the heaving displacement is expressed as:
\begin{equation}
    H^* = \frac{h(t)}{C}, t^* = \frac{tU_\infty}{C}
\end{equation}
where $C$ is the characteristic length (e.g., chord length) and $U_\infty$ is the free-stream velocity. The Bretschneider spectrum is defined in the frequency domain by:
\begin{equation}
    S(f) = \frac{5}{16} H^2_s \frac{f_p^4} {f^{5}} \exp\Big(-\frac{5}{4} \Big(\frac{f_p}{f}\Big)^4\Big)
    \label{Bret_spectra}
\end{equation}
where $S(f)$ is the wave energy density at frequency $f$, $H_s$ is the significant wave height and $f_p$ is the peak frequency. To generate the time series $h(t)$, the spectrum is discretized into a finite number of frequency components, each assigned a random phase, and summed as:
\begin{equation}
    h(t) = \sum_{n=1}^{N} \sqrt{2S(f_n)\Delta f} \cos(2 \pi f_n t + \phi_n)
    \label{ew:Brett_wave-height}
\end{equation}
where $f_n$ are the discrete frequencies, $\Delta f$ is the frequency spacing, and $\phi_n$ are random phases uniformly distributed in $[0, 2\pi]$. 
Fig. \ref{Bret_intro} (b), shows three realizations of the non-dimensional heaving motion generated by irregular wave with non-dimensional significant wave height ($H_s^*=H_s/C$) of 0.62 and $\text{St}_C=f_pC/U_\infty$ of 0.12. The plot also shows a sinusoidal wave with frequency equal to $f_p$ and wave energy equal to that of the irregular wave. On equating the total energy of irregular (Eq. \ref{E_irregular}) and regular wave (Eq. \ref{E_regular}), the non-dimensional wave height of the sinusiodal wave, which is twice its amplitude ($H_0^*$), comes out to be $2H_0^*=H_s^*/\sqrt{2}$.

The present configuration is characterized by six governing parameters $Re$, $I^*$, $X_e^*$, $H_s^*$, St$_C$ and either $f_{\theta}/f_p$ (for the spring-limiter) or $\theta_{0}$ (for the angle-limiter). This set of parameters underscores the intrinsic complexity of the problem. For a systematic comparison between the spring-limiter and the angle-limiter, we fix the values of $Re=10,000$, $I^*=0.28$, $\theta_i=0$, and $X_e^*=0.1$, while varying the spring stiffness (expressed as $f_{\theta}/f_h$), the maximum permissible pitching angle in the angle-limiter ($\theta_{0}$), and the sea-state parameters ($H_s^*$ and St$_C$). Since the simulations are conducted with irregular heaving motions, the influence of these parameters is examined across three independent realizations of the heaving motion. The selected Reynolds number is sufficiently large to sustain coherent vortex shedding and flutter-induced dynamics, while remaining computationally feasible for fully resolved simulations. The considered value of $I^*$ corresponds to solid-to-fluid density ratio of 10. Although variations in $I^*$ could provide further engineering insight, it is kept fixed here to reduce the dimensionality of the parameter space. Nevertheless, it is important to note that $I^*$ directly influences the natural pitching frequency, following the scaling $f_\theta \sim 1/\sqrt{I^*}$, which is pertinent to the present analysis. Finally, the pitch-axis location is prescribed as $X_e^* = 0.1$, consistent with our previous work~\cite{raut2024hydrodynamic}, where this choice was shown to maximize the thrust generated by a single submerged flapping-foil propulsion system.  

In this study, three distinct sea-states are considered, each corresponding to different significant wave heights, with the underlying objective of evaluating the performance of the selected propulsor design across a wide range of ocean conditions. The sea-states are chosen based on annual wave statistics for the North Atlantic, as reported by Wang \emph{et al.} \cite{wang2019dynamic}, and are summarized in Table \ref{tab_sea_states}. As defined in our earlier work \cite{raut2024hydrodynamic}, the Strouhal number based on wake width for regular (monochromatic) wave 
\begin{equation}
\text{St}^\text{reg}_w = {2 h_0 f}/{U_\infty}    
\label{st_reg}
\end{equation}
where $h_0$ represents wave amplitude, $f$ is the wave encounter frequency, and $U_\infty$ is the vehicle speed. In the case of stochastic waves, both the frequency and amplitude of the wave change with time and the definition of the Strouhal number has to be modified to accommodate this feature. For waves based on the Bretschneider spectrum, a straightforward option is to characterize the wave height by the significant wave height $H_s$, and the representative wave frequency by the peak frequency of the energy spectrum $f_p$. Accordingly, we can define wake-width based Strouhal number for the irregular waves as
\begin{equation}
  \text{St}^\text{irr}_w=H_sf_p/U_\infty  
\label{st_irr}
\end{equation}
 whose values are provided in Table \ref{tab_sea_states}. We do not currently know if this definition of the Strouhal number is suitable for scaling the performance of the WAP foil in irregular waves, and we will examine this issue later in the paper. 
 
For the computation of non-dimensional parameters, the foil chord length is fixed at $C = 16$ cm, consistent with the design employed in the ``Wave Glider$^\text{®}$,'' a wave-assisted flapping foil propulsion system developed by Liquid Robotics (a Boeing company) \cite{yang2018numerical}. Furthermore, based on operational data for the Wave Glider, the corresponding vehicle speeds are taken to be 0.8 kt, 1.4 kt, and 1.4 kt for sea-states 1, 2, and 3, respectively.
\begin{table}[h!]
\centering
\begin{tabular}{>{\centering\arraybackslash}p{2.2cm}>{\centering\arraybackslash}p{2.2cm}>{\centering\arraybackslash}p{2.2cm}>{\centering\arraybackslash}p{1.5cm}>{\centering\arraybackslash}p{1.5cm}>{\centering\arraybackslash}p{1.5cm}>{\centering\arraybackslash}p{1.5cm}>{\centering\arraybackslash}p{2.5cm}}
\hline
\hline
 Sea-state \# & Wave height (m) &  Wave period (s) &  $H_s^*$ & $H_0^*$ & St$_C$ & $\text{St}^\text{reg}_w $ & $\text{St}^\text{irr}_w $ \\
 \hline
 1 & 0.1 & 3.3 & 0.62 & 0.22 & 0.12 & 0.05 & 0.07 \\ 
 2 & 0.5 & 5   & 3.12 & 1.1 & 0.04 & 0.08 & 0.12\\
 3 & 0.88 & 7.5 & 5.5 & 1.94 & 0.03 & 0.12 & 0.17\\ 
 \hline
\end{tabular}
\caption{Details of the sea-states used in the present study. The regular and irregular waves have the same energy for each sea-state. Here, $H_s^*$ represents significant wave height of the irregular wave, $H_0^*$ represents amplitude of the regular wave, St$_C$ represent chord-based and peak frequency-based Strouhal number and St$_w$ represent wake-width based Strouhal number (Eq. \ref{st_reg} and \ref{st_irr}).}
\label{tab_sea_states}
\end{table}

\subsection{Computational Method}
The fluid-structure interaction in this study is modeled using the sharp-interface immersed boundary method solver \textbf{ViCar3D}, as detailed by Mittal \emph{et al.} \cite{mittal2008versatile} and Seo \& Mittal \cite{seo2011sharp}. This approach is particularly well-suited for simulations involving moving and deforming solid geometries, as it employs a Cartesian grid that does not conform to the body but accurately represents its shape and motion. The method preserves sharp interfaces around the immersed boundary, enabling precise computation of surface quantities \cite{mittal2008versatile}. The governing flow equations are solved using a second-order fractional-step scheme, while the pressure Poisson equation is handled via a bi-conjugate gradient solver. Spatial derivatives are discretized with second-order finite differences, and time integration combines the second-order Adams-Bashforth and Crank-Nicolson schemes. The solver \textbf{ViCar3D} has been extensively validated for both stationary and moving boundary problems in prior studies \cite{seo2023hydrodynamics, prakhar2025bioinspired, kumar2025computational}.

A loosely coupled strategy is employed to model the fluid-structure interaction, wherein the flow and structural equations are solved sequentially. Hydrodynamic forces and moments are computed at Lagrangian marker points distributed on the surface of the solid body and subsequently applied to the structural dynamics equations (Eq. \ref{solid_equation}). The resulting angular velocities are then imposed back on the marker points. A two-dimensional hydrodynamic representation is adopted for the WAP foils in this study. Previous simulations \citep{raut2024hydrodynamic, raut2025dynamics, raut2025harnessing} have shown that intrinsic three-dimensional effects have negligible influence on the thrust generation of such foils. Therefore, the current 2D simulations provide a reasonable approximation for large-span WAP foils, such as those employed in the Wave Glider, which features a foil aspect ratio of 6.3 \cite{kraus2012wave}.

The hydroelastic system is simulated within a computational domain of size $10C \times 15C$, with the leading hydrofoil positioned 4.5 chord lengths downstream of the upstream boundary. The grid near the foil is isotropic, containing approximately 250 points along the chord, and grid stretching is applied in all directions outside the region surrounding the foil and near wake, resulting in a baseline grid of $641 \times 1676$ points. At the inlet, a Dirichlet velocity boundary condition is imposed, while zero-gradient Neumann conditions are applied at all other boundaries. Previous grid refinement and domain-dependence studies \cite{raut2024hydrodynamic, raut2025dynamics} confirm that this resolution yields well-converged, domain-independent results.

In all simulations, the foils are initially placed at zero pitch angle at $y=0$ (the mean heaving position), with zero initial angular and heaving velocities. The equilibrium angle of attack is also set to zero. A constant free-stream flow is applied, allowing the system dynamics to evolve naturally over a duration of eight times the wave period $T_p$ corresponding to the peak frequency $f_p$. For all cases, mean quantities are computed over the last four periods $T_p$, ensuring statistically steady behavior.
%
\section{Results} 
%
\subsection{\label{regular_irregular_comp}Foil Kinematics for Regular and Irregular Waves}
\begin{table}
\begin{tabular}{>{\centering\arraybackslash}b{2.7cm}  >{\centering\arraybackslash}b{5.9cm} >{\centering\arraybackslash}b{5.2cm} } 
 \hline
 \hline
 Sea-state \# & $\theta_0$ (deg.) (angle-limiter) &  $f_\theta/f_p$ (spring-limiter) \\
 \hline
 1 & 0, 1.25, 2.5, 3.75, 5, 7.5 &  2, 3, 4, 6,  8 \\ 
 2 & 2.5, 5, 7.5, 10, 12.5  &  6, 9, 12, 15, 18   \\
 3 & 7.5, 10, 12.5, 15, 17.5 &  6, 9, 12, 15, 18, 21 \\ 
 \hline
\end{tabular}
\caption{Parameters chosen for angle-limiter and spring-limiter to study pitch limiting mechanism.}
\label{tab_params}
\end{table}
For the flow-induced motion simulations of the WAP system, an irregular heaving motion is prescribed for the hydrofoil depending on the sea-state. The hydrofoil is allowed to pitch passively based on the pitching moment imposed on it by the flow and the spring/angle-limiter. Table \ref{tab_params} summarizes the parameters chosen for {\color{black}the} set of simulations that are carried out for the study of the pitch constraining mechanisms. Thus, a total of 48 different irregular wave simulations each for the spring-limiter and for {\color{black}the} angle-limiter are being performed, comprising 16 distinct parameter variations with three independent simulations/realizations performed for each variation, to gain a comprehensive understanding of the effect of this mechanism on the thrust performance. Along with the irregular wave, regular or sinusoidal wave simulations are performed for the 16 distinct parameter variations with the heaving amplitude chosen such that the total wave energy is equal to the irregular wave at different sea-states.

\begin{figure}
    \centering
    \subfigure[  ]{\includegraphics[width=0.475\textwidth, trim={0cm 0cm 1cm 0cm}, clip]{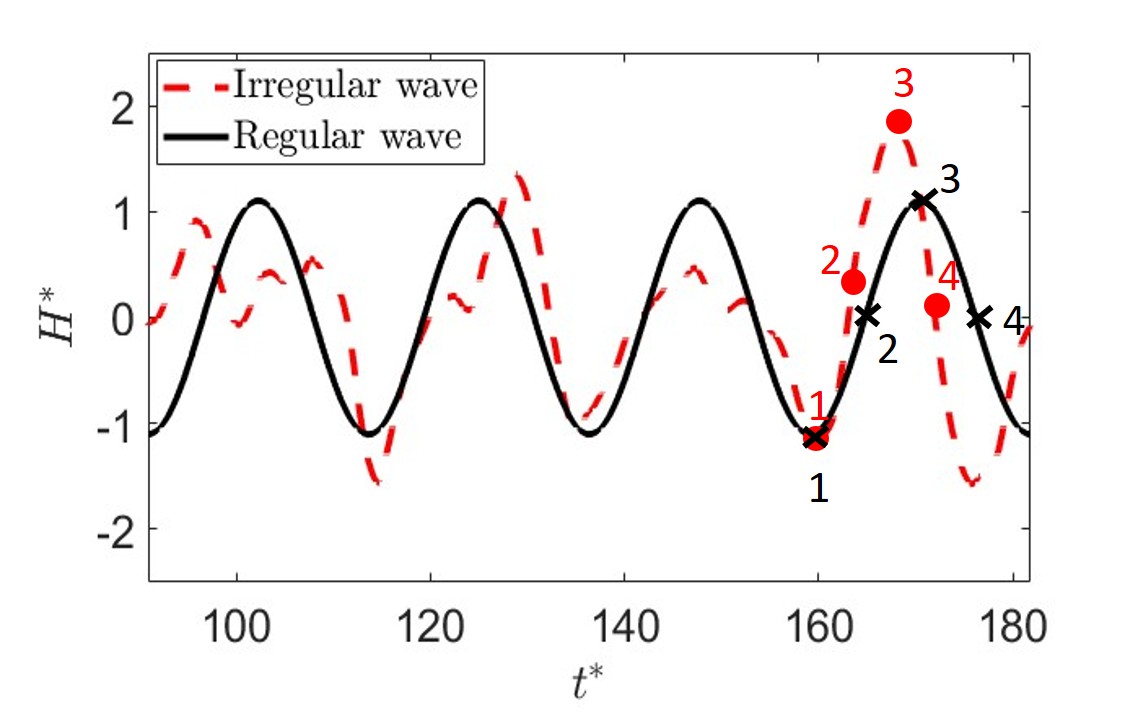}
    \label{compare_heav}}
    \subfigure[ ]{\includegraphics[width=0.5\textwidth,height=0.225\textheight, trim={0cm 0cm 1cm 0cm}, clip]{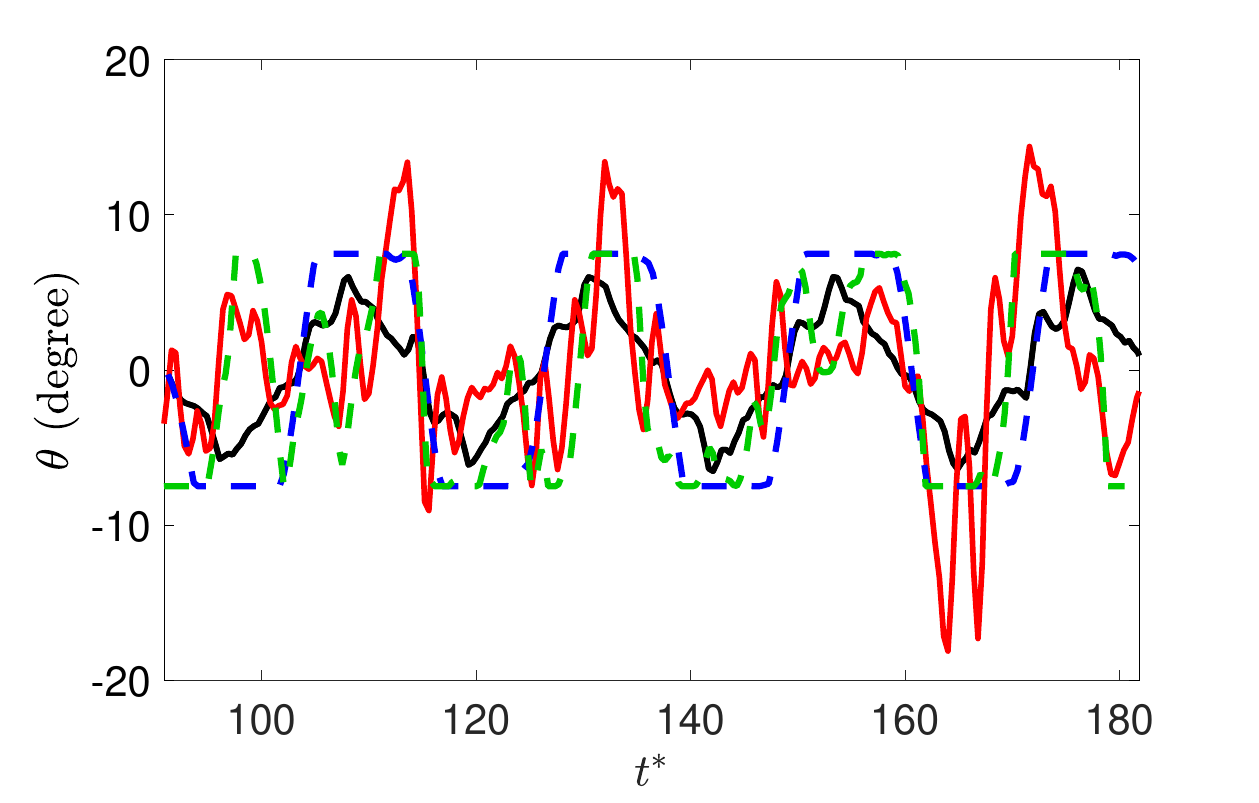}
    \label{compare_posi}}\\
    \subfigure[ ]{\includegraphics[width=0.5\textwidth, trim={0cm 0cm 0cm 0.0cm}, clip]{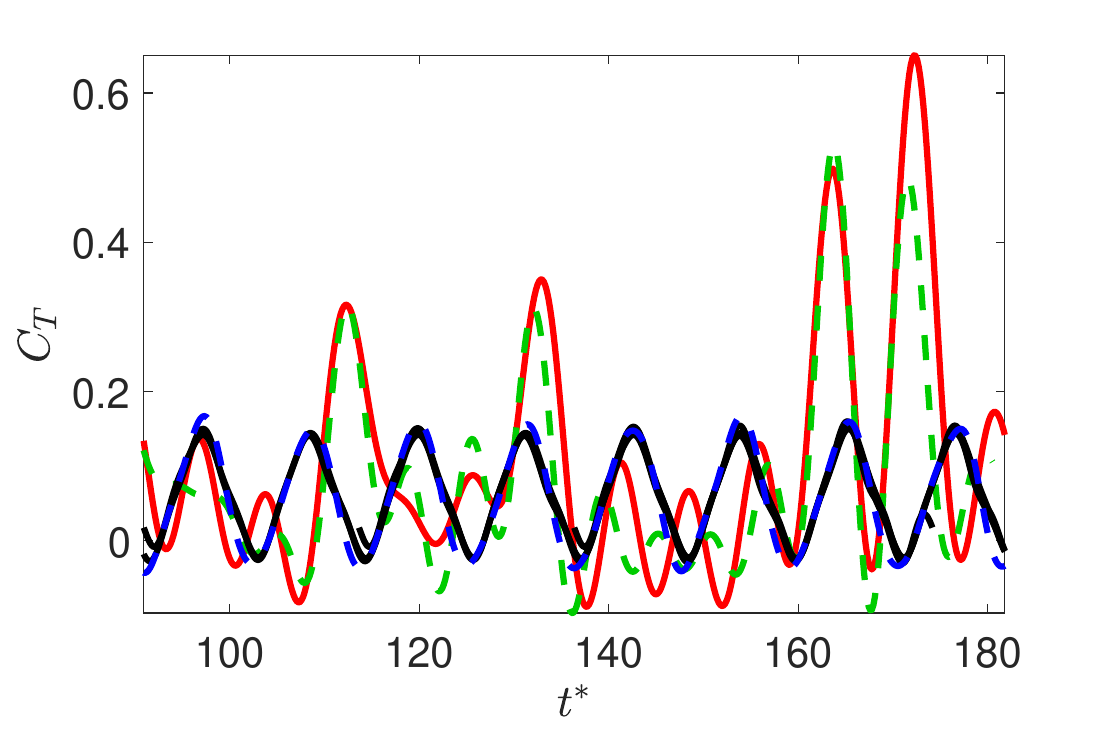}
    \label{compare_thrust}}
    \subfigure[]{\includegraphics[width=0.485\textwidth, height=0.245\textheight, trim={0cm 0cm 0cm 0.0cm}, clip]
    {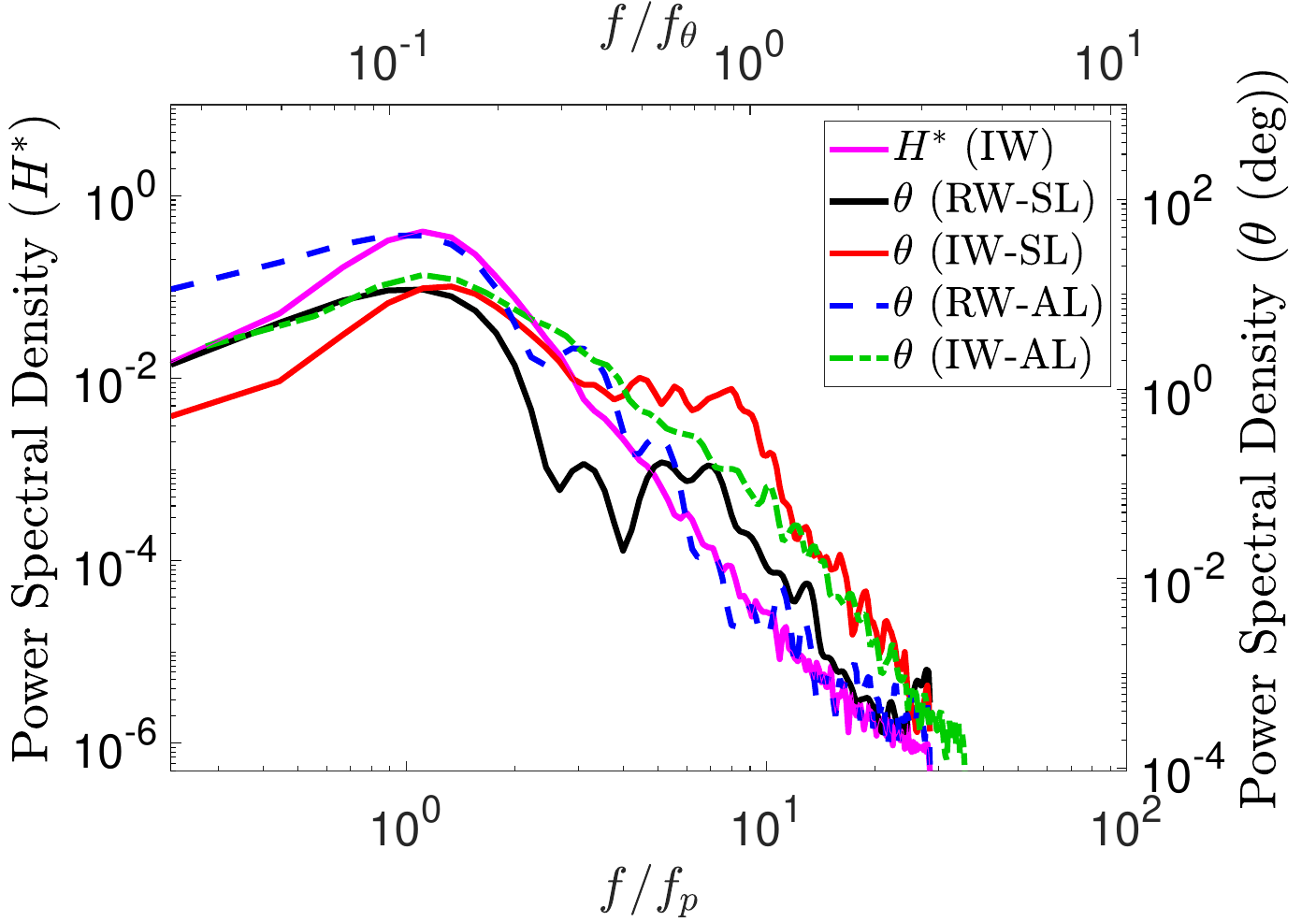}\label{heave_pitch_spec}}
    \caption{Comparison between the regular and the irregular wave conditions for the (a) heaving motion, (b) pitching motion and (c) thrust coefficient for sea-state 2 with spring-limiter ($f_\theta/f_h=9$) and angle-limiter ($\theta_0=7.5$) as pitch constraining mechanism. (d) shows the power spectral density for the heaving and pitching motion in regular and irregular wave condition corresponding to (a) and (b). (IW: Irregular Wave, RW: Regular Wave, SL: Spring Limiter and AL: Angle Limiter)}
    \label{}
    \vspace{1cm}
\end{figure}
We begin by presenting in Fig. \ref{compare_heav}, the comparison of heaving motion for the foil under regular and irregular wave condition with same total energy under sea-state 2 (Table \ref{tab_sea_states}). Only one realization of the irregular wave is shown here for comparison. Sea-state 2 is selected for demonstrating the foil kinematics, as it represents an intermediate case between sea-states 1 and 3 in terms of the Strouhal number, St$_w$. The observed trends for this case are consistent with those obtained for the other sea-states. Owing to the stochastic character of the irregular waves, the foil experiences rapid transient heaving motions, such as that evident within the interval $t^*=160$ to $t^*=180$. Fig. \ref{compare_posi}, shows the comparison between the pitching motion for the foil under regular and irregular wave condition with both spring and angle-limiter as the pitch-constraining mechanism. The acronyms RW-SL, IW-SL, RW-AL and IW-AL used in the figures represent Regular wave (RW) or Irregular wave (IW) with spring-limiter (SL) or angle-limiter (AL) and have been used for rest of the paper. A frequency ratio $f_\theta/f_h=9$ is selected for the spring-limiter in this initial description, as it is subsequently demonstrated (Section \ref{compare_AL_SL}) that this value yields enhanced thrust generation under irregular wave conditions corresponding to sea-state 2. A pitch amplitude ($\theta_0$) of $7.5^\circ$ is chosen for the angle-limiter because it generates a rms (root mean square) of pitch oscillation $\theta_\text{rms}$ of $6.04^{\circ}$ which is close to the rms of pitch oscillation for {\color{black}the} spring-limiter case ($\theta_\text{rms}=5.74^\circ$) under irregular wave condition. The spring-limiter exhibits a complex pitch response that consists of rapid variations driven by the elastic recoil of the torsional spring. The sharp heaving motion under irregular wave condition leads to sharp pitching motion and this can be seen more notably within the interval $t^*=160$ to $t^*=180$. 

Fig. \ref{compare_thrust}, shows the time series for the coefficient of thrust ($C_T=-F_x/\frac{1}{2}\rho U^2_\infty C$) within the same time interval as in Fig. \ref{compare_heav} and \ref{compare_posi}. This shows that the intensified heaving and pitching motion due to the stochastic nature of the irregular wave leads to a sharp temporal rise in the thrust coefficient, which is again most notable in the interval $t^*=160$ to $t^*=180$. This enhancement in thrust for the foil could be due to enhancement in leading-edge vortex of the foil or due to acceleration-reaction effects. This will be discussed more in section \ref{forcepartitioning}, where we apply the force-partitioning method (Appendix \ref{FPM_section}) to decompose the pressure-induced forces on the foil. 

Figure \ref{heave_pitch_spec} shows the power spectrum of heaving and pitching motion for regular and irregular waves. The spectrum of the pitching motion for the irregular waves is more broadband as expected with substantial power at the higher frequencies. This is associated with the rapid variations that we see in the pitching motion under irregular waves in Fig. \ref{compare_posi}. The angle-limiter has a higher peak as compared to spring-limiter at $f/f_p=1$ (bottom x-axis) since the angle-limiter does not have a spring to restrict it's motion. On the other hand, for the spring limiter, there is a band of frequencies ($f/f_p$) between 5 and 10 where there is increased power and this is associated with a resonance response with the torsional spring which has a natural frequency set at $f_\theta/f_p=9$.

\begin{figure}
    \centering
    \vspace{-0.5cm}
    \subfigure[  ]{\includegraphics[width=0.29\textwidth, trim={0cm 0cm 0cm 0cm}, clip]{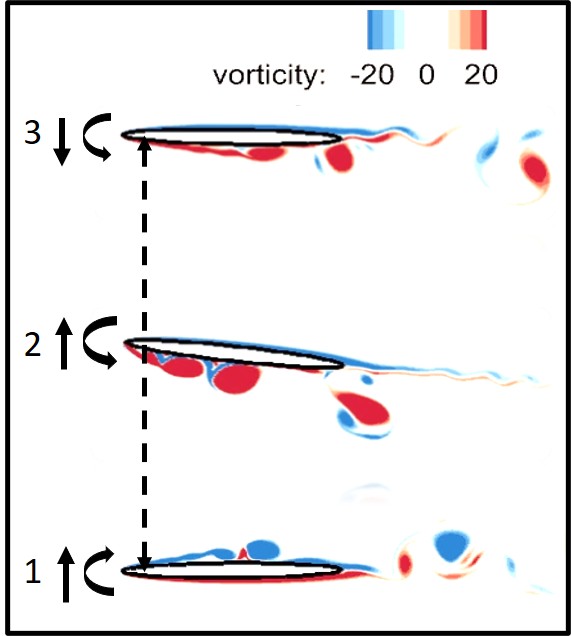}
    \label{regular_springL}}
    \hspace{1cm}
    \subfigure[ ]{\includegraphics[width=0.47\textwidth, trim={0cm 0cm 0cm 0cm}, clip]{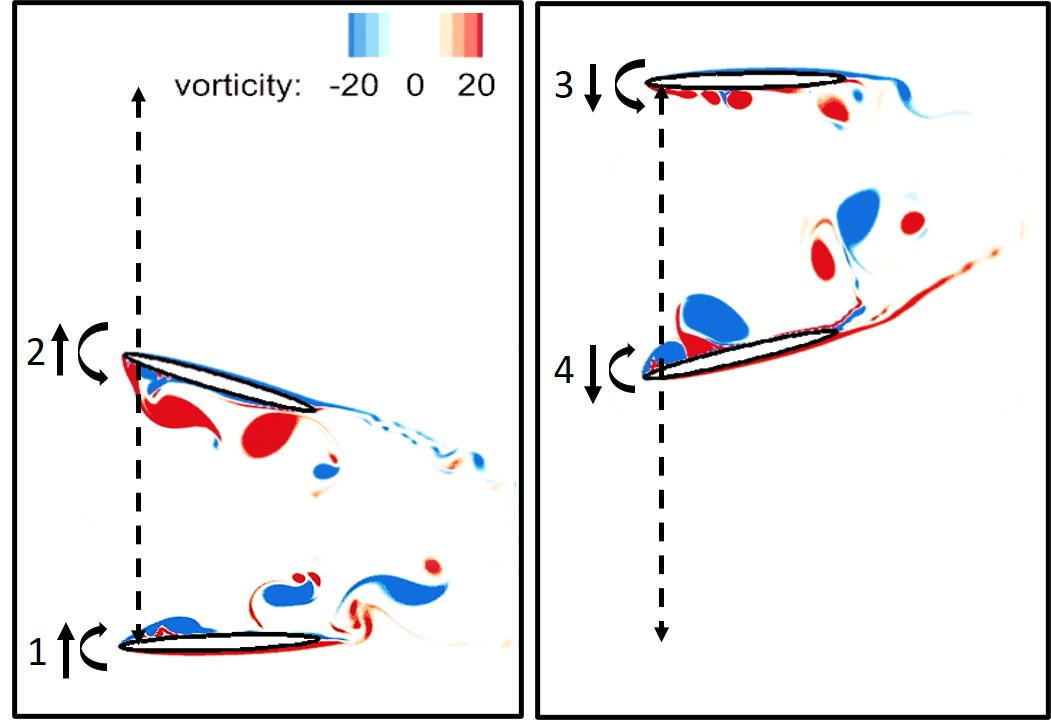}
    \label{irregular_springL}}\\
    \subfigure[ ]{\includegraphics[width=0.47\textwidth, trim={0cm 0cm 0cm 0.0cm}, clip]{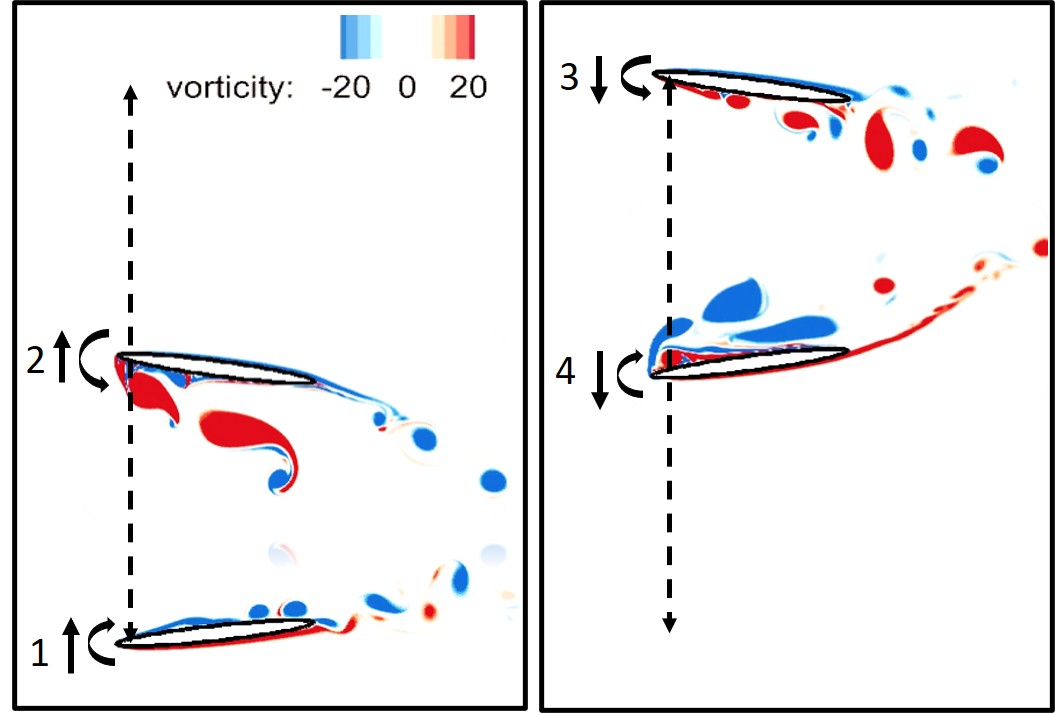}
    \label{irregular_angleL}}
    \caption{Vorticity contour plots for (a) regular wave with spring-limiter ($f_\theta/f_h=9$), (b) irregular wave with spring-limiter ($f_\theta/f_h=9$) and (c) irregular wave with angle-limiter ($\theta_0=7.5^\circ$) for sea-state 2.}
    \vspace{1cm}
\end{figure}

Fig \ref{regular_springL} and \ref{irregular_springL} shows the vortex shedding for the foil under regular and irregular wave condition respectively, with spring-limiter as pitch-constraint mechanism and on the points marked within the interval $t^*=160$ to $t^*=180$ in Fig. \ref{compare_heav}. For the regular wave, as the hydrofoil starts to heave in the upward direction, it experiences an increasing clockwise (negative) moment resulting in the pitching up of the foil. However, this pitch-up and heave-up motion increases the effective angle-of-attack of the foil and leads to the formation of leading-edge vortices (LEVs)  on the suction surface of the hydrofoil as seen in Fig. \ref{regular_springL}(2). These LEVs generate a suction pressure on the adjacent surface and this generates a positive thrust and a negative lift force. During the downstroke, the LEV formation process switches to the top surface resulting in {\color{black}the} generation of a positive lift and a positive thrust. A similar process of LEV formation occurs in the case of the irregular wave, as can be seen in Fig. \ref{irregular_springL} and \ref{irregular_angleL} for spring-limiter and angle-limiter respectively. 
However, for temporal instances (2) and (4) which correspond to the large heaving velocities, the foil with the spring-limiter can be seen to exhibit a higher pitch angle than the foil with the angle-limiter, and this is due to the intrinsic mechanism of the angle-limiter which limits the pitch angle to a certain values irrespective of the heave velocity. We will show later that this is key to the better performance of the spring limiter in irregular wave conditions.
%
%

\subsection{\label{compare_AL_SL}Performance of Angle-limiter Versus Spring-limiter for Different Sea-States}
We next perform a comparative study between the spring-limiter and the angle-limiter designs in terms of thrust generation by varying the stiffness (or $f_\theta$) of the torsional spring and the maximum angle allowed by the angle-limiter ($\theta_{0}$) for different sea-states shown in table \ref{tab_sea_states}. Due to the stochastic nature of the irregular wave, three different realizations are considered for each sea-state as shown in Fig. \ref{Bret_intro} (b) for sea-state 1. Simulations for the regular wave with the same total energy as irregular wave are also conducted to study the impact of irregularity of the surface wave on thrust generation for each sea-state. 

\begin{figure}
    \centering
    \vspace{-0.5cm}
    \subfigure[]{
    \includegraphics[height=0.25\textheight,width=0.55\textwidth, trim={0.1cm 0.0cm 2.3cm 1.8cm}, clip]{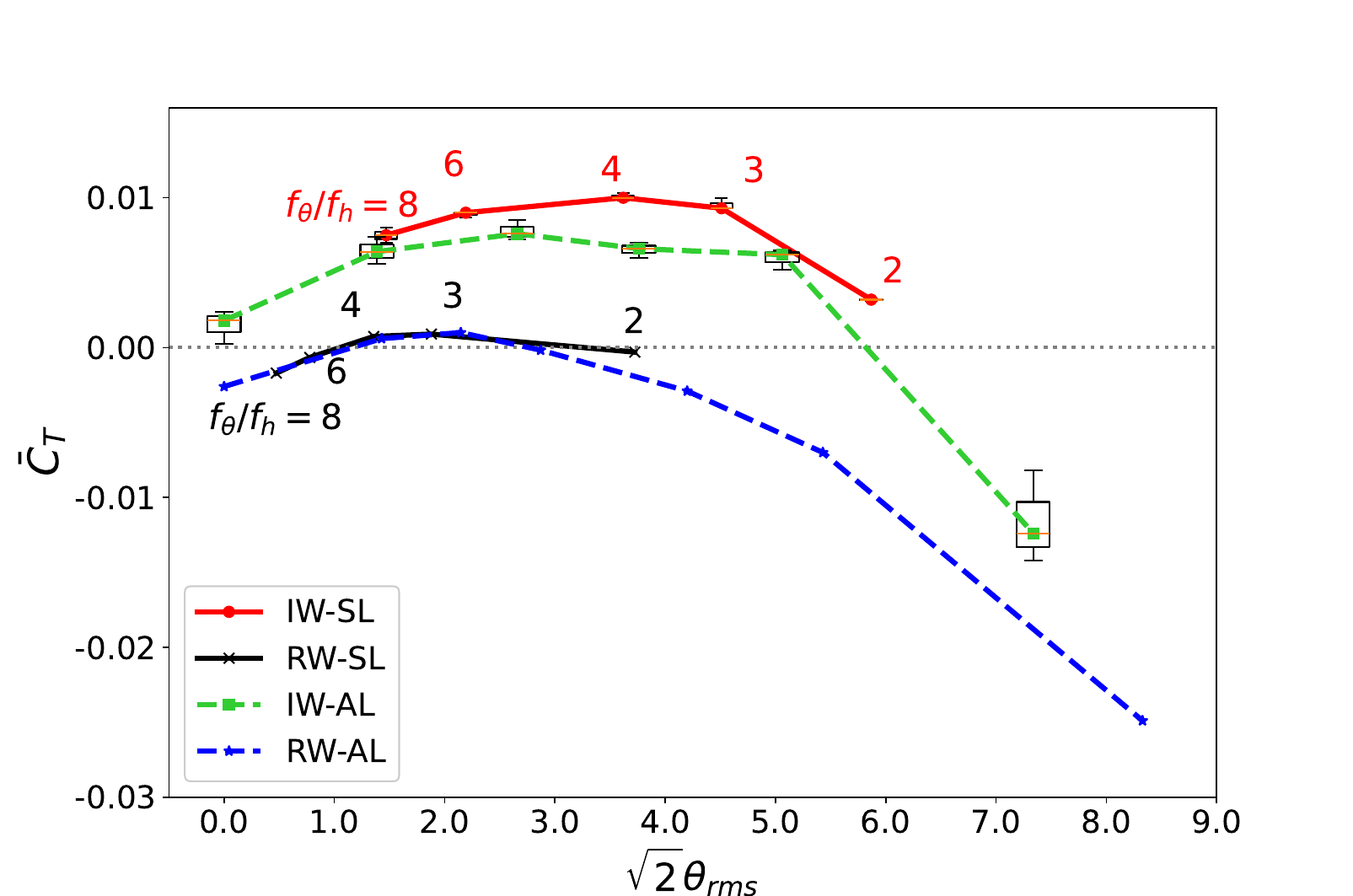}}\\
    \subfigure[]{
    \includegraphics[height=0.25\textheight,width=0.55\textwidth, trim={0.1cm 0.0cm 2.3cm 1.7cm}, clip]{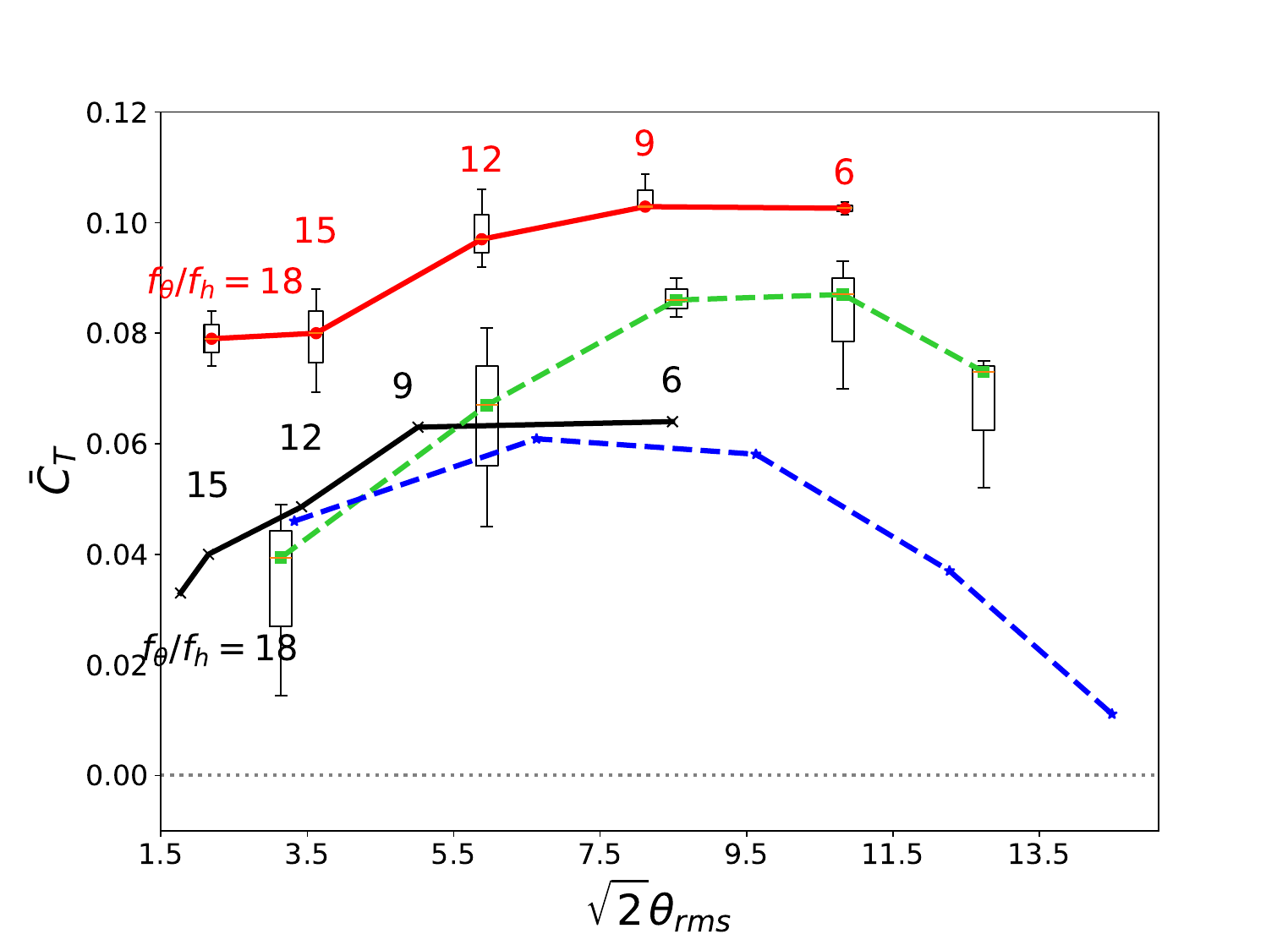}}    
    \subfigure[]{
    \includegraphics[height=0.25\textheight,width=0.55\textwidth, trim={0.1cm 0.0cm 2.3cm 1.8cm}, clip]{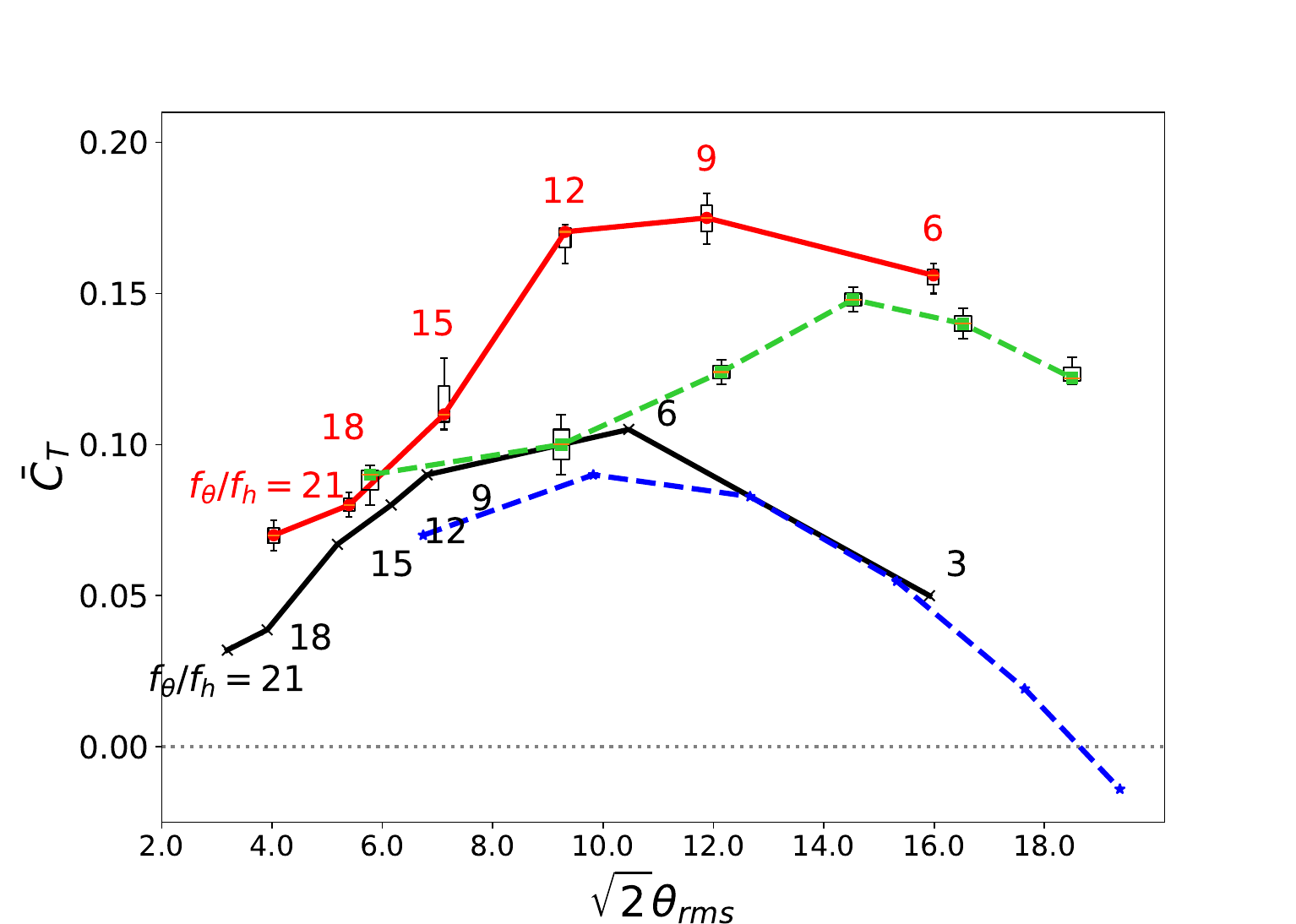}}    \caption{Average thrust coefficient ($\overline{C}_T$) over one oscillation cycle for angle-limiter at different pitching amplitudes for (a) sea-state 1 ($H_0^*=0.31, \text{St}_C=0.12$), (b) sea-state 2 ($H_0^*=1.56, \text{St}_C=0.04$) and (c) sea-state 3 ($H_0^*=2.75, \text{St}_C=0.03$)}
    \label{ss1_thrust}
    \vspace{1cm}
\end{figure}

Fig. \ref{ss1_thrust} (a) shows the average thrust generated by the spring-limiter ($\bar{C}_T$) in sea-state 1 for different frequency ratio ($f_\theta/f_h$) which is equivalent to changing the stiffness of the torsional spring. Given that simulations for three realizations are performed for the irregular wave case, we represent the thrust generated using a bar and whisker plot based on the results from these three simulations for each $f_\theta/f_h$. The results are plotted against the root-mean-square value of pitching motion ($\theta_\text{rms}$), allowing for a consistent comparison for all the cases. Sea-state 1 is the most challenging sea-state for a WAP propulsor since the wave-height is small and indeed, we find that only a small thrust is generated for this sea-state for all the values of $f_\theta/f_h$ that we have employed. The highest thrust is generated for $f_\theta/f_h=4$ which has a $\theta_\text{rms}$ of 2.56$^\circ$. 

Fig. \ref{ss1_thrust} (a) also shows the same results for the angle-limiter with varying pitch amplitude ($\theta_0$), leading to different values of $\theta_\text{rms}$ on $x$-axis. This mechanism generates lower thrust as compared to the spring-limiter with the highest value occurring at $\theta_\text{rms}$ of 1.88$^\circ$ (or $\theta_0=2.5^\circ$). The maximum thrust generated by the spring-limiter has been found to be nearly 29\% higher than the maximum thrust generated by the angle-limiter. Fig. \ref{ss1_thrust} (a) also shows the thrust generated by the spring-limiter and angle-limiter under regular wave conditions with varying $f_\theta/f_h$ and $\theta_0$ respectively. For most of the values of $f_\theta/f_h$ and $\theta_0$, both the spring-limiter and angle-limiter generate drag under regular wave conditions. On comparing the peak values, the angle-limiter performs slightly better (by 10\%) than the spring-limiter, which was also shown in our earlier work \cite{raut2025harnessing} for small wave heights.
\begin{figure}
    \centering
    \vspace{-0.5cm}
    \includegraphics[height=0.25\textheight,width=0.55\textwidth, trim={0.1cm 0.0cm 2.3cm 1.0cm}, clip]{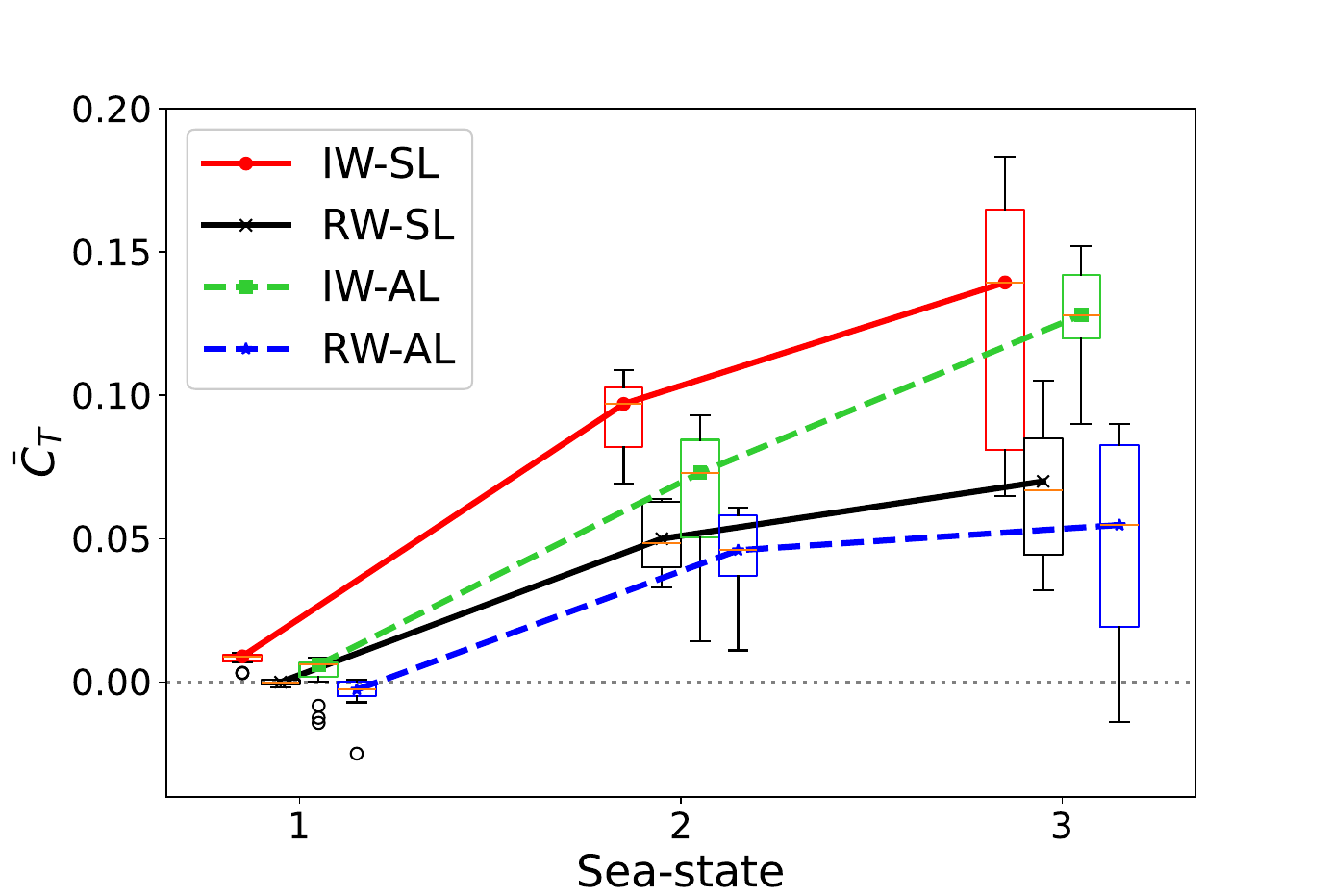}\\   
    \caption{Bar and whisker plot for the average thrust coefficient for the regular and irregular waves with spring-limiter and angle-limiter as pitch-constraining mechanisms.}
    \label{whisker_allstates}
    \vspace{1cm}
\end{figure}

Fig. \ref{ss1_thrust} (b) shows the average thrust coefficient for sea-state 2 using both the angle-limiter and the spring-limiter mechanisms under regular and irregular wave conditions. Compared to sea-state 1, the optimum value for pitch amplitude has shifted towards right with the optimum value for the angle-limiter occurring at $\theta_\text{rms}=7.65^\circ$ ($\theta_0 = 10^\circ$) and for {\color{black}the} spring-limiter at $\theta_\text{rms} = 5.74^\circ$ ($f_\theta/f_h=9$). The maximum thrust from spring-limiter has been found to be 18\% higher than the maximum thrust generated with the angle-limiter. Similarly for sea-state 3 (Fig. \ref{ss1_thrust} (c)), the optimum value of $\theta_\text{rms}$ for the angle-limiter and the spring-limiter shifts towards right and now occur at $\theta_\text{rms}=10.29^\circ$ ($\theta_0=12.5^\circ$) and $\theta_\text{rms}=8.42^\circ$ ($f_\theta/f_h=9$) respectively. For this sea-state, the maximum thrust from the spring-limiter has been found to be 18\% higher than the maximum thrust generated from the use of angle-limiter. The thrust generated from the equivalent energy regular wave can be seen to be significantly lower than the irregular waves for both sea-states, with spring-limiter giving slightly better thrust performance than angle-limiter.

Figure~\ref{whisker_allstates} presents a summary of the overall thrust performance, expressed in terms of the average thrust, for all the conditions examined in this study. For regular wave cases, there is minimal difference in the performance between the spring-limiter and angle-limiter configurations, particularly at lower sea states. In sea-state~1, the peak thrust observed for the angle-limiter case is higher than that of the spring-limiter case, whereas reverse is true for sea-state 2 and 3. This observation is consistent with our previous findings on the influence of pitch-constraining mechanisms in regular wave conditions~\cite{raut2025harnessing}. Under irregular wave conditions, the WAP system generates substantially higher thrust compared to regular waves. Moreover, for all irregular wave cases, the spring-limiter configuration outperforms the angle-limiter by a significant margin across all sea states. The subsequent sections of this paper further investigate and elucidate the underlying reasons for the observed differences in performance among these cases.

\subsection{\label{forcepartitioning} Partitioning of the Thrust Forces on the Foil}
\begin{figure}
    \centering
    \vspace{-0.5cm}
    \subfigure[]{
    \includegraphics[width=0.47\textwidth, trim={0cm 0cm 0cm 0cm}, clip]{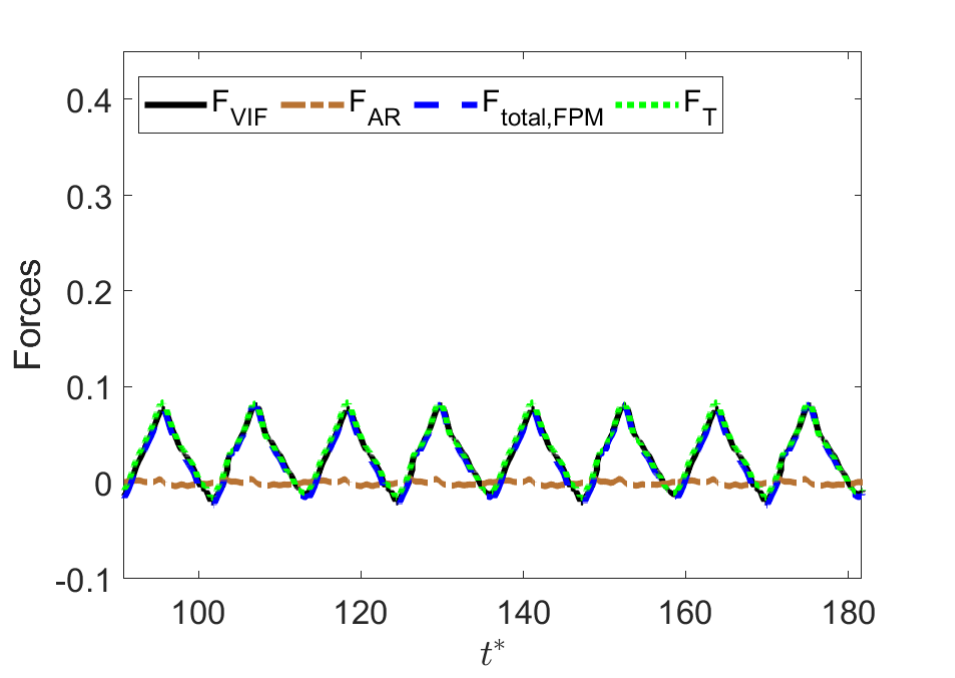}}
    \subfigure[]{
    \includegraphics[width=0.47\textwidth, trim={0cm 0cm 0cm 0cm}, clip]{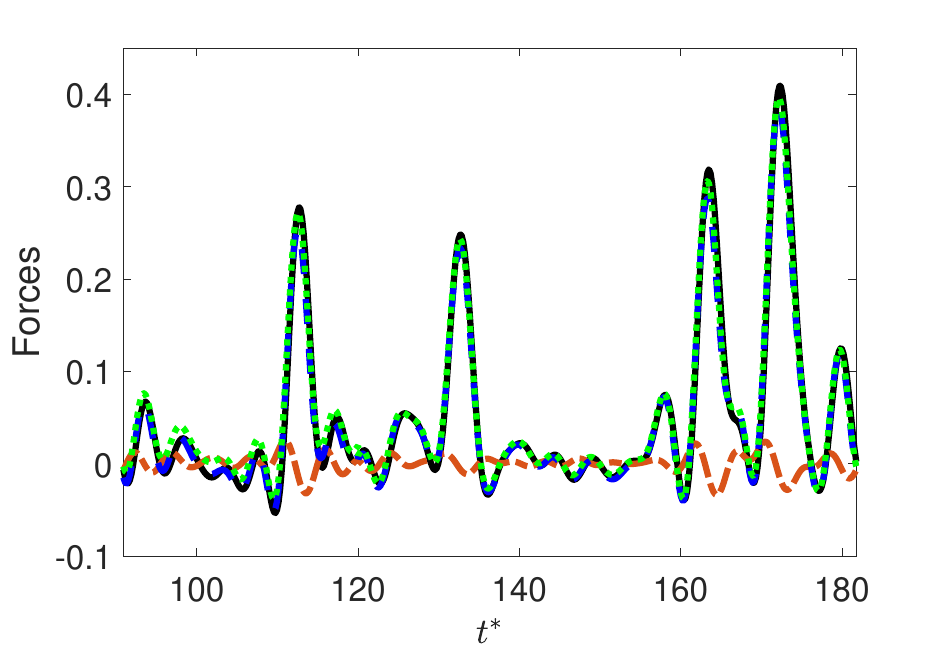}}\\
    \subfigure[]{
    \includegraphics[width=0.47\textwidth, height=0.23\textheight, trim={0.0cm 0cm 0cm 0cm}, clip]{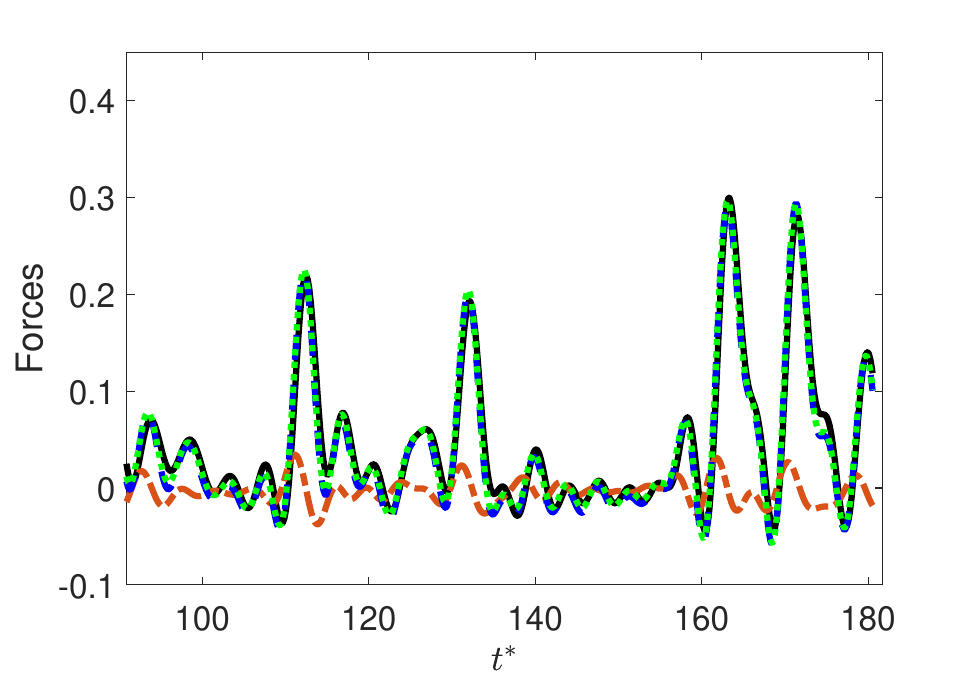}}
    \caption{Partitioned pressure-induced forces in $x$-direction into individual components using the force partitioning method applied to (a) regular wave with spring-limiter (b) irregular wave with spring-limiter ($f_\theta/f_h=9$) and (c) irregular wave with angle-limiter ($\theta_0=7.5^\circ$) for sea-state 2.} 
    \label{fpm_applied}
    \vspace{1cm}
\end{figure}

We begin the analysis by applying the force partitioning method (FPM) (with details in Appendix \ref{FPM_section}) to the forces in the thrust ($-x$) direction. FPM is a method that allows us to determine the dominant mechanisms for force generation on a body immersed in a fluid \citep{zhang2015centripetal,menon2021quantitative,raut2024hydrodynamic}. Fig. \ref{fpm_applied} shows FPM applied to three cases with one involving a regular wave with spring-limiter ($f_\theta/f_p=9$) and the other two involving irregular waves with spring- ($f_\theta/f_p=9$) and angle-limiters ($\theta_0=7.5^\circ$). The two non-negligible components are the vortex-induced force $\text{F}_\text{VIF}$ and force induced due to acceleration reaction $\text{F}_\text{AR}$ (also identified as added-mass effects) and these have been plotted in the figure. 
The values of $\text{F}_\text{VIS}$ and $\text{F}_\text{BOUND}$ has been found to be very small for all cases in Fig. \ref{fpm_applied} and hence not shown. 
$\text{F}_\text{AR}$ has small but non-zero instantaneous values but it turns out that the average thrust from $\text{F}_\text{AR}$ is negligible. Thus, nearly the entire mean thrust is associated with $\text{F}_\text{VIF}$. 

\subsection{Thrust analysis using the LEV-based model (LEVBM)}
We have found in our previous work \cite{raut2024hydrodynamic,raut2025dynamics} that it is the leading-edge vortex (LEV) that is responsible for generating most of the vortex-induced thrust. We have also proposed and tested a LEV based model (LEVBM) to predict the thrust from such flapping foil system\cite{raut2024hydrodynamic,zhou2025hydrodynamically}. We now use this model to further understand the difference in the performance of the foil in sinusoidal and irregular waves, as well as to understand why the choice of the pitch-limiting mechanism also leads to differing thrust for each of these wave conditions.
\begin{figure}[h!]
    \centering
    \subfigure[]{\includegraphics[width=0.48\textwidth, trim={0cm 0cm 0cm 0.0cm}, clip]{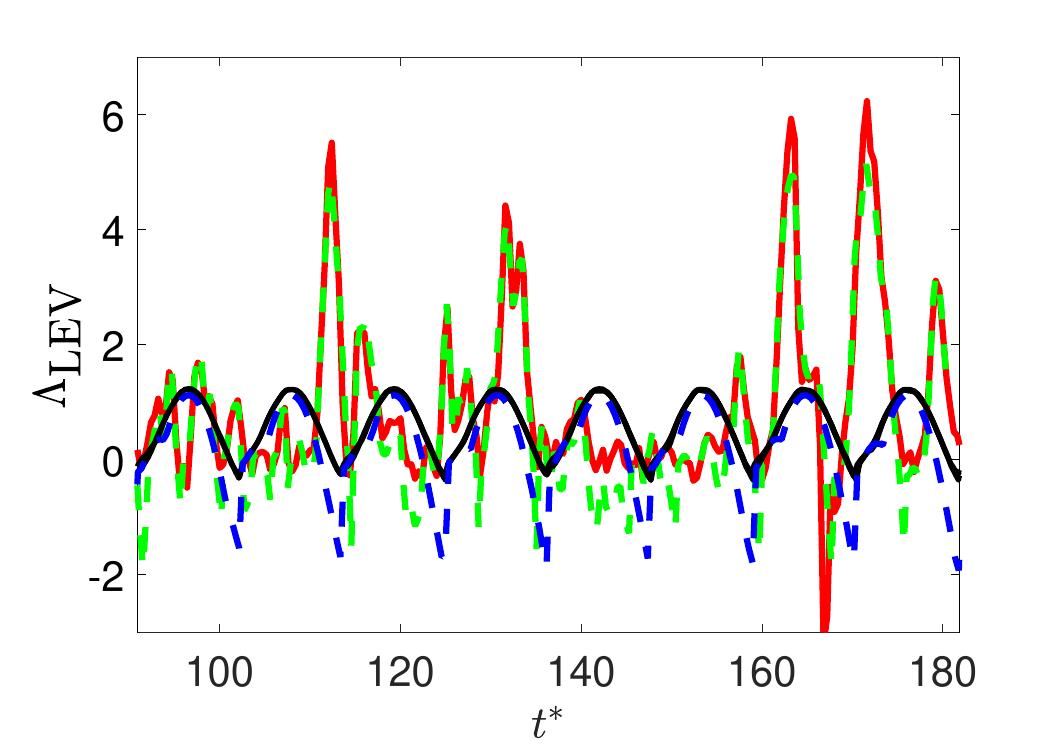}}
    \subfigure[  ]{\includegraphics[width=0.485\textwidth, trim={0cm 0cm 0cm 0cm}, clip]{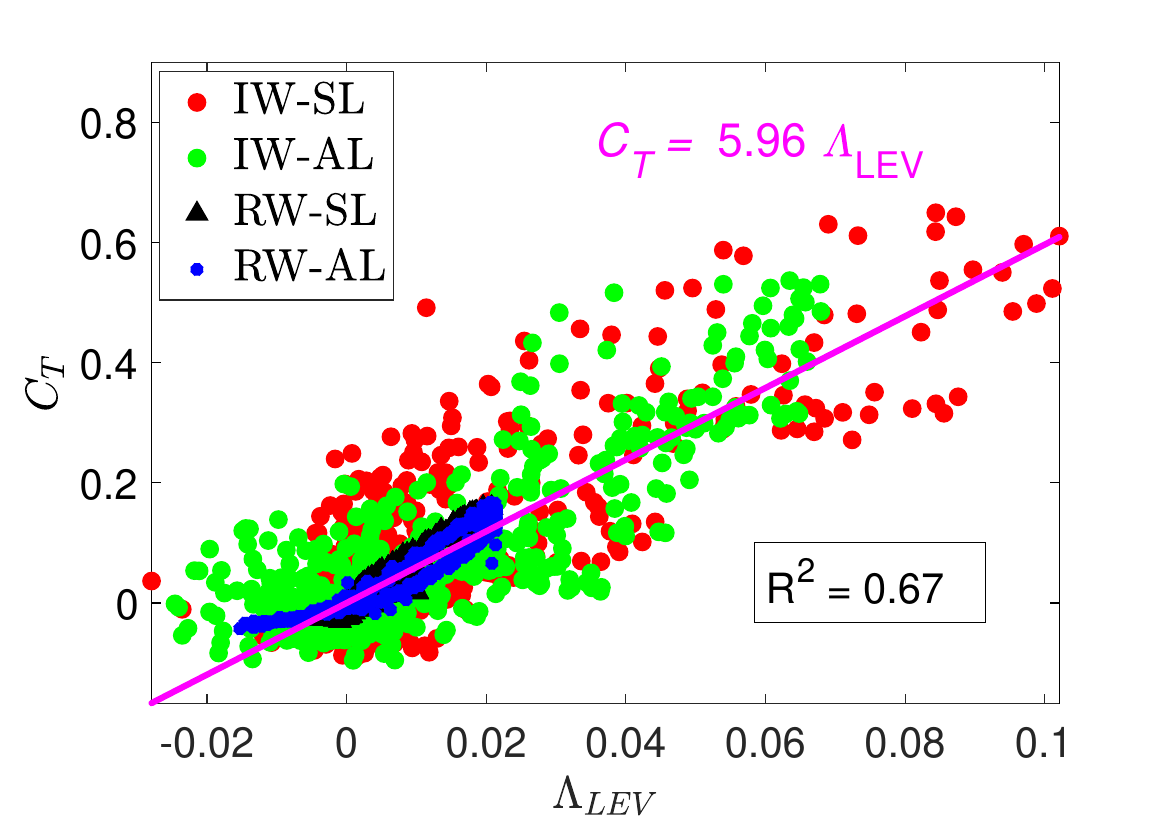}
    \label{}}
    \caption{(a) Time series for $\Lambda_{\rm LEV}$ calculated based on the kinematics of the foil (Eq. \ref{kappa_eq}) for sea-state 2 with regular and irregular wave conditions and with spring-limiter ($f_\theta/f_h=9$) and angle-limiter ($\theta_0=7.5^\circ$) as pitch-constraint mechanism. (b) Scatter plot between $C_T$ and $\Lambda_{\rm LEV}$ based on the time series plots in Fig. \ref{compare_thrust} and (a).}
    \label{LEVBM_compare}
    \vspace{1cm}
\end{figure}
As per the model, the strength of the LEV is proportional to the $\sin{(\alpha_\text{eff})}$ where $\alpha_\text{eff}$ is the effective angle-of-attack given by 
\begin{equation}
\alpha_\text{eff}(t)= \tan^{-1}\left(-\dot{h}(t)/U\right) -\theta(t) 
\label{eq:alpha_eff}
\end{equation}
where $\dot{h}$ is the heave velocity of the hinge-point and $\theta$ is the pitch angle of the foil. Subsequently, the LEV-induced thrust can be expressed as 
\begin{equation}
    C_T (t) = K \, \Lambda_\textrm{LEV}(t)
    \label{eq_kappa2}
\end{equation}
where $K$ is proportionality constant and 
\begin{equation}
   \Lambda_\textrm{LEV}(t)=\sin{(\alpha_\text{eff})} \sin{\theta}.
   \label{kappa_eq}
\end{equation}
In the above expression $C_T$ is the thrust coefficient expressed as 
\begin{equation}
    C_T = \frac{F_T}{\frac{1}{2}\rho V_\textrm{max}^2 C}
    \label{eq:C_T}
\end{equation}
here $F_T$ is the thrust force on the foil and $V_\textrm{max}$ is given by $V_\textrm{max}=\sqrt{U^2_\infty+(h_0 2\pi f_h)^2}$. Here, $f_h$ can be calculated from St$_C$ as $f_h = \text{St}_C U_\infty/C$. A detailed derivation of Eq. \ref{eq_kappa2} is shown in Appendix \ref{LEVBM_model} along with the validation originally shown in our earlier paper. 

Fig. \ref{LEVBM_compare} shows the evolution of $\Lambda_{\rm LEV}$ for the kinematics shown in Fig. \ref{compare_posi}. This time series matches closely with the thrust coefficient plot in Fig. \ref{compare_thrust}, confirming a direct relationship between the two for both regular and irregular waves. This is further validated in the scatter plot in Fig. \ref{LEVBM_compare} (b) which shows a linear relationship between $C_T$ and $\Lambda_{\rm LEV}$ for the time series values in Fig. \ref{compare_thrust} and Fig. \ref{LEVBM_compare} (a). This shows that the model works for high values of $\Lambda_{\rm LEV}$ and the predictions are valid outside the range of sinusoidally flapping foils. 

We now examine and explain (a) why the large transient heaving motions that appear in irregular wave motions are able to generate large values of $\Lambda_{\rm LEV}$, and consequently of thrust and (b) why the spring-limiter performance is better than the angle-limiter. 

\subsection{Coherence between Heaving and Pitching}
\begin{figure}
    \centering
    \subfigure[  ]{\includegraphics[width=0.45\textwidth, trim={0cm 0cm 0cm 0cm}, clip]{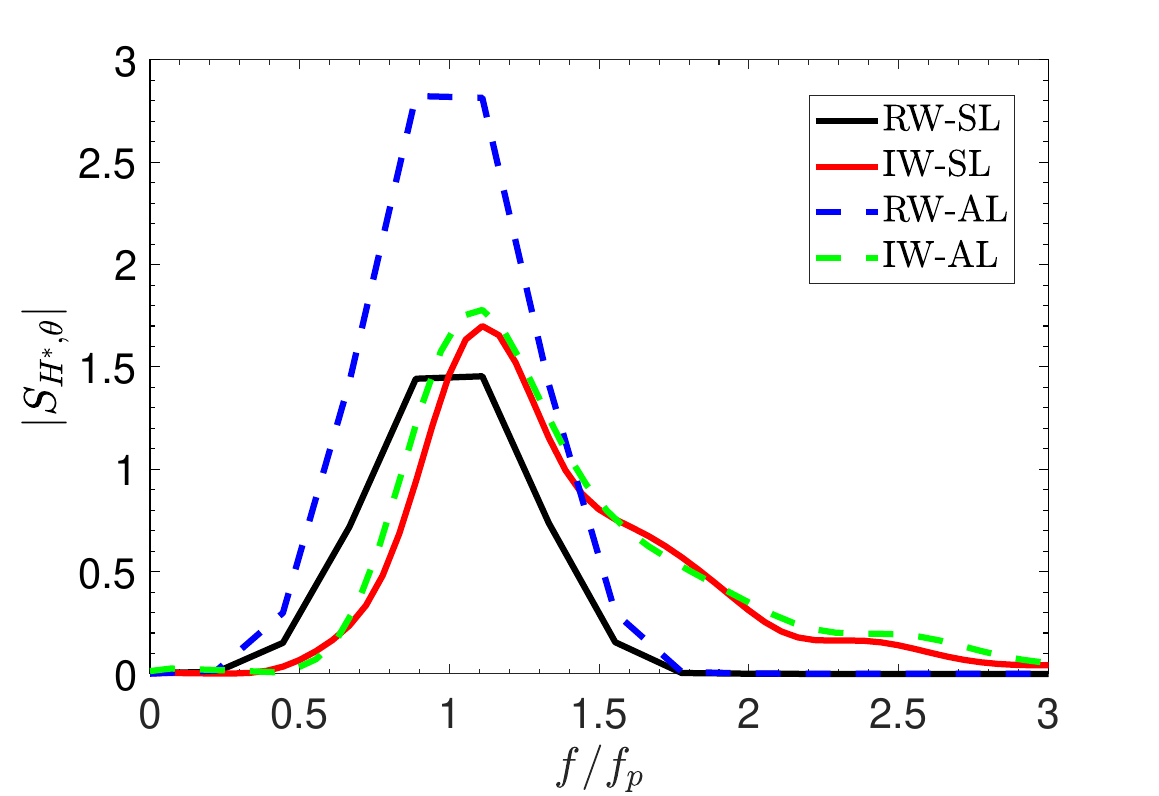}}
    \subfigure[  ]{\includegraphics[width=0.45\textwidth, trim={0cm 0cm 0cm 0cm}, clip]{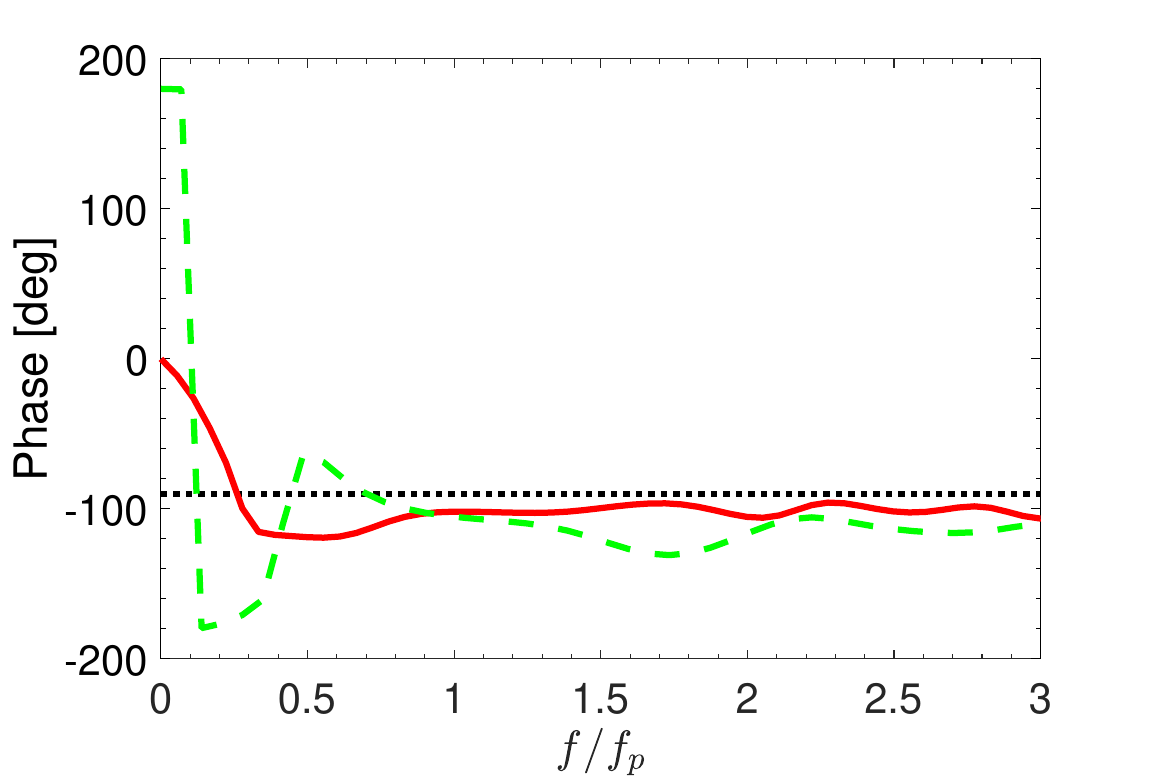}}
        \subfigure[  ]{\includegraphics[width=0.45\textwidth, trim={0cm 0cm 0cm 0cm}, clip]{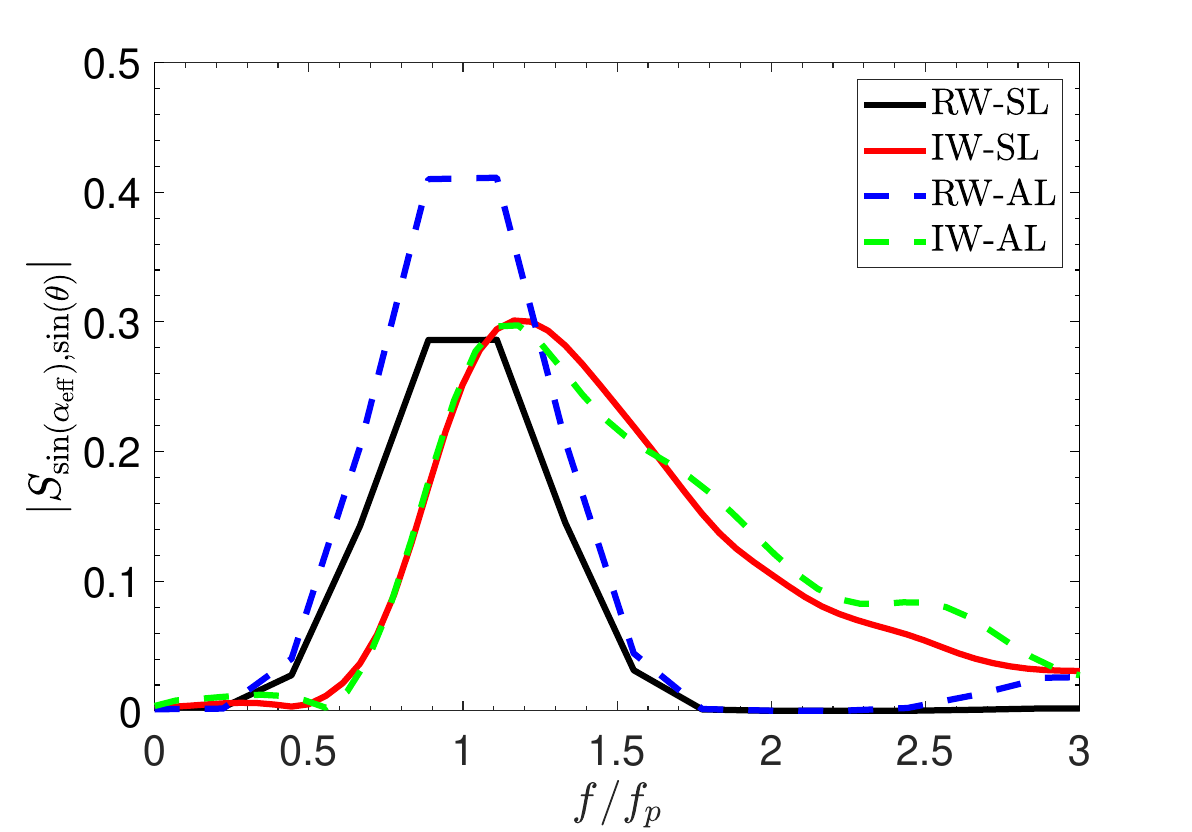}}
    \subfigure[  ]{\includegraphics[width=0.43\textwidth, height=0.228\textheight, trim={0cm 0cm 0cm 0cm}, clip]{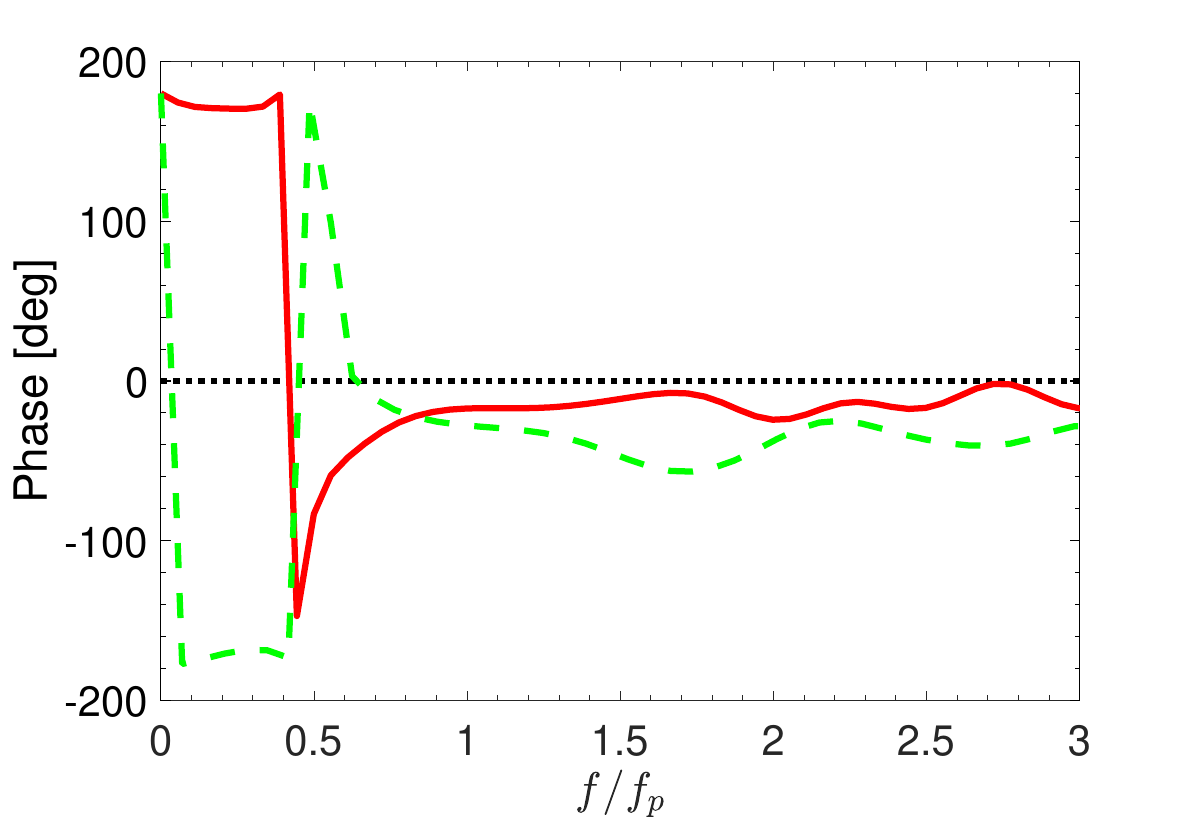}}
        \caption{(a) Cross spectral density magnitude and (b) phase of the cross-spectrum between heaving ($H^*$) and pitching motion ($\theta$). (c) Cross spectral density magnitude and (d) phase of the cross-spectrum between $\sin(\alpha_{eff})$ and $\sin(\theta)$. All the plots are for sea-state 2 with spring-limiter ($f_\theta/f_h=9$) and angle-limiter ($\theta_0=7.5$) in regular and irregular wave conditions.}
    \label{csd_heave_pitch}
    \vspace{1cm}
\end{figure}

As shown in our previous work\cite{raut2024hydrodynamic} and also evident from \ref{kappa_eq}, to maximize $\Lambda_{\rm LEV}$, $-\dot{h}$ has to be in phase with $\theta$. This can be achieved with a pitching-heaving phase difference of $-\pi/2$. To quantify the frequency-domain relationship between heave and pitch motion, we compute the cross-spectral density (CSD) between their respective time series as shown in Fig. \ref{csd_heave_pitch} (a). The CSD provides a measure of how strongly two signals are correlated as a function of frequency. In this context, it allows us to identify the dominant frequencies at which the foil’s motion variables, i.e. heave and pitch, exchange energy, as well as the phase differences between them. Peaks in the CSD magnitude indicate frequencies where both signals are highly coherent with each other, while the corresponding phase spectrum reveals phase lead or lags between the signals. This information is crucial for understanding how coupled kinematic parameters contribute to unsteady hydrodynamic force generation. In this analysis, we present the ensemble average of the cross-spectrum and phase over all three realizations for each case. 

As can be seen in Fig. \ref{csd_heave_pitch}(a), there exists a strong coherence between the heave and pitch motion for frequencies close to the peak frequency $f_p$ for all the cases. For the periodic waves, we note that the cross-spectrum has higher peak values for the angle-limiter when compared to the spring-limiter and this is associated with the restoring force due to the spring that would tend to resist foil pitching and which also generates a recoil at the ends of the heaving stroke. For the irregular waves, the cross-spectrum for the two pitch limiting mechanisms is quite similar indicating that for these cases, the irregularity of the waves result in weakening the role of the pitch-limiting mechanisms on the performance. 

Fig. \ref{csd_heave_pitch} (b) shows the phase difference between the two signals as a function of frequency for spring-limiter and angle-limiter in irregular wave condition. We see that the phase difference is close to $-\pi/2$, especially in case of spring-limiter, and this should lead to higher values of $\Lambda_{\rm LEV}$ and subsequently higher thrust. Fig. \ref{csd_heave_pitch} (c) shows the CSD between $\sin(\alpha_{\rm eff})$ and $\sin(\theta)$ which are the two terms involved in calculation of $\Lambda_{\rm LEV}$ (Eq. \ref{kappa_eq}). Here again we see strong correlation between the two signals close to the peak frequency $f_p$ for all the cases. Fig. \ref{csd_heave_pitch} (d) shows the phase difference between the two signals as a function of frequency for the spring-limiter and the angle-limiter in irregular wave condition. The phase difference is close to zero, especially in the case of the spring-limiter, which directly correlates to a better thrust performance. 

\subsection{Enhanced Performance of Spring-Limiter in Irregular Waves\label{sec:irr_wave_compare}}
The coherence analysis clarifies that the spring-limiter configuration enables a more favorable pitching response compared to the angle-limiter. Under high-amplitude heave motions, the spring-limiter allows correspondingly large pitching amplitudes (Fig.~\ref{compare_posi}) thereby leading to high $\Lambda_\text{LEV}$ and thrust. On the other hand, for the angle-limiter case, the pitch angle remains capped by the mechanical stopper, thereby limiting the values of $\Lambda_\text{LEV}$. This restriction also prevents the system from attaining the favorable phase difference of $-\pi/2$ between the heave and pitch motions (Fig.~\ref{csd_heave_pitch}(b)) during the rapid heaving events characteristic of irregular waves. In contrast, the compliance introduced by the spring-limiter naturally supports this desirable phase relationship, thereby enhancing thrust generation. By imposing a rigid constraint on the pitch motion, the angle-limiter disrupts this coupling, ultimately leading to reduced overall performance.

\subsection{Enhanced Performance of WAP in Irregular Waves}
In this final section, we provide an explanation for the enhanced thrust generated by the irregular wave motion when compared to the sinusoidal wave motion despite an equivalence in the total wave power between the two conditions. For this, we turn to the LEVBM model based prediction of thrust versus Strouhal number for optimal pitching conditions. In particular LEVBM indicates that for any given Strouhal number, there is a pitch amplitude ($\theta_0$) that maximizes thrust (a detailed derivation of the equation is in \citep{raut2024hydrodynamic} and a succinct description is given in appendix \ref{LEVBM_model}), which is given by 
\begin{equation}
    \theta^\text{opt}_{0} = 0.5 \tan^{-1}\left( \pi \text{St}_w \right)-\theta_s
    \label{eq:theta0_max_eq_LEVBM}
\end{equation}
The parameter $\theta_s$, is a small angle-offset which is related to the thickness and shape of the foil at the leading-edge, and this can be neglected for the current thin elliptic foil. Based on the optimal value of pitch amplitude $\theta^{\text{opt}}_0$, the optimal value of thrust can be calculated by substituting this optimal value into Eqs. \ref{eq:alpha_eff} - \ref{eq:C_T} 
and the average value of thrust coefficient can be computed by taking a mean over one cycle.
\begin{figure}
    \centering
    \vspace{-0.5cm}
\includegraphics[height=0.27\textheight,width=0.6\textwidth, trim={0.1cm 0.0cm 0cm 0cm}, clip]{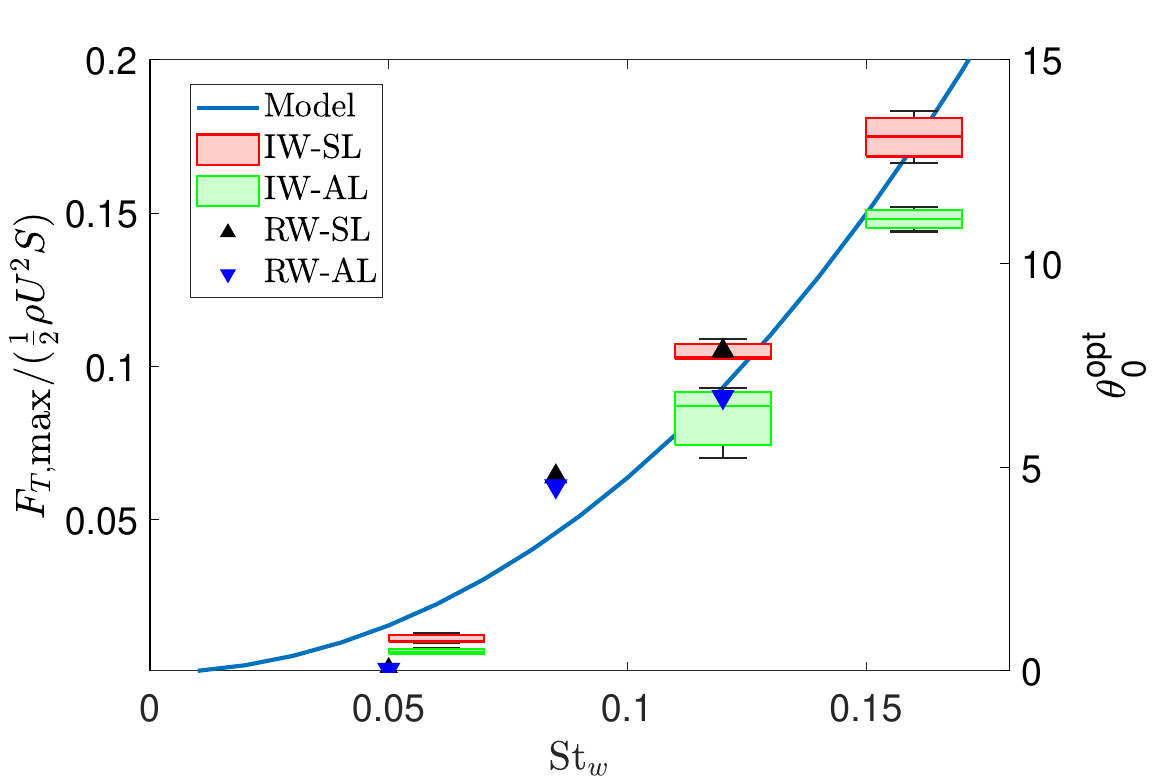}\\   
    \caption{Comparison between LEVBM model and DNS for the optimal thrust. We use St$_w=2fh_o/U_{\infty}$ for regular waves and St$_w=\sqrt{2}\dot{h}_\text{rms}/(\pi U_{\infty})$ for irregular waves.}
    \label{model_DNS_comp}
    \vspace{1cm}
\end{figure}
%


The above formula was derived for sinusoidal waves for which $\text{St}_w$ was defined as $2fh_o/U_{\infty}$. The symbols in the Fig. \ref{model_DNS_comp} correspond to the DNS results for the sinusoidal waves and they exhibit a reasonable match to the scaling law in Eq. \ref{eq_kappa2}, confirming the validity of the scaling law. 
For irregular waves, there is no characteristic frequency and heaving amplitude that can be used to define the Strouhal number and earlier in the paper, we used $\text{St}_{w}^{\text{irr}}=H_sf_p/U_\infty $ as a possible definition of the Stouhal number for the irregular waves. However, a more nuanced and flow-physics based definition of Strouhal number is required for correct scaling, and noting that $2\pi fh_o$ is in fact a measure of the maximum heave velocity, we define the Strouhal number for the irregular waves based on a measure of the maximum heave velocity as 
\begin{equation}
\text{St}_{w}^{\text{irr}^*} = \sqrt{2}\dot{h}_\text{rms}/(\pi U_{\infty})
\label{st_irr2}
\end{equation}
where $\dot{h}_\text{rms}$ is the ensemble averaged root-mean-square of the measured heave velocity of the foil in irregular waves. 


The above definition of the Strouhal number allows us to plot the results for the irregular waves alongside those for the sinusoidal waves. This comparison in Fig. \ref{model_DNS_comp} shows that firstly, plotted in this way, the computed variation of thrust for the irregular waves also tracks the predictions from the LEVBM, thereby confirming the validity of the formula, as well as the use of heave velocity as a characteristic variable in this formula. Second, the plot shows that for equal wave energy, the effective Strouhal number of the irregular waves is higher than the corresponding value for the sinusoidal waves at each sea-state. Indeed, this difference gets very large for the higher sea-states - the Strouhal number for the irregular waves at seas-state 2 is nearly the same as that for the sinusoidal waves at sea-state 3, and for sea-state 3, the Strouhal number for the irregular waves is 41.67\% higher than that for the sinusoidal waves. It is this larger effective Strouhal number, combined with the constructive coherence and phasing between the pitching and heaving, that results in the significantly better thrust performance of the WAP systems in irregular waves. 

We note in passing that, for the Bretschneider spectrum, $
\left( \frac{H_s f_p}{U_\infty} \right) 
= 1.005 \left( \frac{\sqrt{2}\,\dot{h}_\mathrm{rms}}{\pi U_\infty} \right).
$
Consequently, the Strouhal number defined for irregular waves earlier in the paper 
(\(\text{St}_{w,\mathrm{irr}}\), Eq.~\ref{st_irr}) is nearly identical to the Strouhal number derived 
from flow-physics-based considerations (\(\text{St}_{w,\mathrm{irr}}^*\), Eq.~\ref{st_irr2}). 
This agreement, however, is purely coincidental and arises from the specific shape of the 
Bretschneider spectrum; it would, for example, not hold for other spectral forms such as 
the JONSWAP spectrum. The flow-physics based Strouhal number is therefore the one that should be used for problems involving irregular (or regular) flapping motion.

\subsection{Guidance for WAP Design}
Overall, the present study demonstrates that the WAP system achieves superior performance under irregular wave conditions with 63.6\% and 66.5\% higher thrusts achieved for sea-states 2 and 3 respectively, compared to sinusoidal waves with the same overall wave energy. Among the two pitch constraining mechanisms examined, the spring-limiter consistently outperforms the angle-limiter across all investigated sea-states, with an improvement in thrust of approximately $20\%$. An ideal implementation of the spring-limiter would incorporate an adaptive stiffness mechanism that adjusts the spring constant in a feedforward manner, informed by an estimate of the Strouhal number (see Fig.~\ref{schematic}(a)). Such an approach could allow the WAP system to maximize thrust for each sea-state. However, this strategy necessitates sensing for wave frequency, wave amplitude, and vessel speed, as well as an actuation system capable of real-time stiffness adjustment, which may increase the overall system complexity. If such complexity is deemed impractical, a simpler configuration employing an angle-limiter with a fixed pitch amplitude may provide a viable alternative. The present findings also offer design guidance for such systems. Specifically, the simulations suggest that for the angle-limiter, setting the pitch amplitude to approximately $\theta_0 = 5^\circ$ is an effective strategy, as this configuration produces thrust across all sea-states with realistic, irregular motion, including sea-state-1, where propulsion generation from wave motion is particularly difficult.

\section{\label{conc}Conclusions} 
There has been limited attention given to the performance of wave-assisted propulsion (WAP) systems operating under irregular wave conditions. In this study, high-fidelity flow simulations incorporating fully coupled fluid--structure interaction modeling have been employed to investigate the hydrodynamic behavior and performance of flapping-foil propulsors subjected to irregular waves. The primary objective of the present work is to elucidate the flow physics governing flapping-foil propulsion under irregular wave excitation and to understand the hydrodynamics of the resulting performance against equivalent regular-wave conditions. 

A direct comparison between regular and irregular waves of equal energy reveals that the WAP system achieves higher efficiency in irregular wave environments. The rapid, large-amplitude heave motions induced by these irregular waves generate sharp peaks in the instantaneous thrust coefficient, resulting in an overall increase in mean thrust. Interpreted through the LEV-based model \cite{raut2024hydrodynamic}, this enhancement can be attributed to the higher wave-based Strouhal number experienced under irregular forcing, since thrust scales monotonically with Strouhal number. This outcome is noteworthy because stochastic forcing is often associated with performance degradation. In the case of WAP systems, however, it represents a fortuitous advantage, as these systems will almost always operate in naturally stochastic wave conditions.

Building on previous work that focused on regular (sinusoidal waves) \cite{raut2025dynamics}, the present study also explores the effect of pitch-constraining mechanisms when operating in irregular waves. Two such mechanisms are examined: a spring-limiter and an angle-limiter, evaluated under three representative sea-states. The results demonstrate that the spring-limiter consistently outperforms the angle-limiter by approximately $20\%$, with particularly pronounced relative gains in lower sea-states, which correspond to  conditions critical for WAP-based propulsion. The underlying cause is that the compliance introduced by the torsional spring enables a more favorable phase relationship between the heaving and pitching motions. In contrast, the rigid constraint imposed by the angle-limiter disrupts this optimal phase synchronization, leading to reduced performance.

In summary, this study demonstrates how a fundamental understanding of the flow physics underlying flapping foil systems can inform the design of more efficient WAP systems. The results highlight the critical role of structural compliance in enabling these systems to adapt to the complexity of unsteady and stochastic wave environments. This principle likely extends to natural and bioinspired flyers and swimmers, including birds, insects, fish, and engineered aerial or underwater vehicles that exploit flexural compliance to maintain stable and efficient locomotion in stochastic environments such as those associated with gusty or turbulent flows.

\subsection*{Acknowledgements}
This work is supported by ONR Grants N00014-22-1-2655 and N00014-22-1-2770. This work used the computational resources at the Advanced Research Computing at Hopkins (ARCH) core facility (rockfish.jhu.edu), which is supported by the AFOSR DURIP Grant FA9550-21-1-0303, and the Extreme Science
and Engineering Discovery Environment (XSEDE), which is supported by National Science Foundation Grant No. ACI-1548562, through allocation number TG-CTS100002.

\subsection*{Declaration of interests}
The authors report no conflict of interest.

\appendix

\section{\label{FPM_section} Force-Partitioning Method}
\begin{figure}
    \centering
    \includegraphics[width=0.5\textwidth]{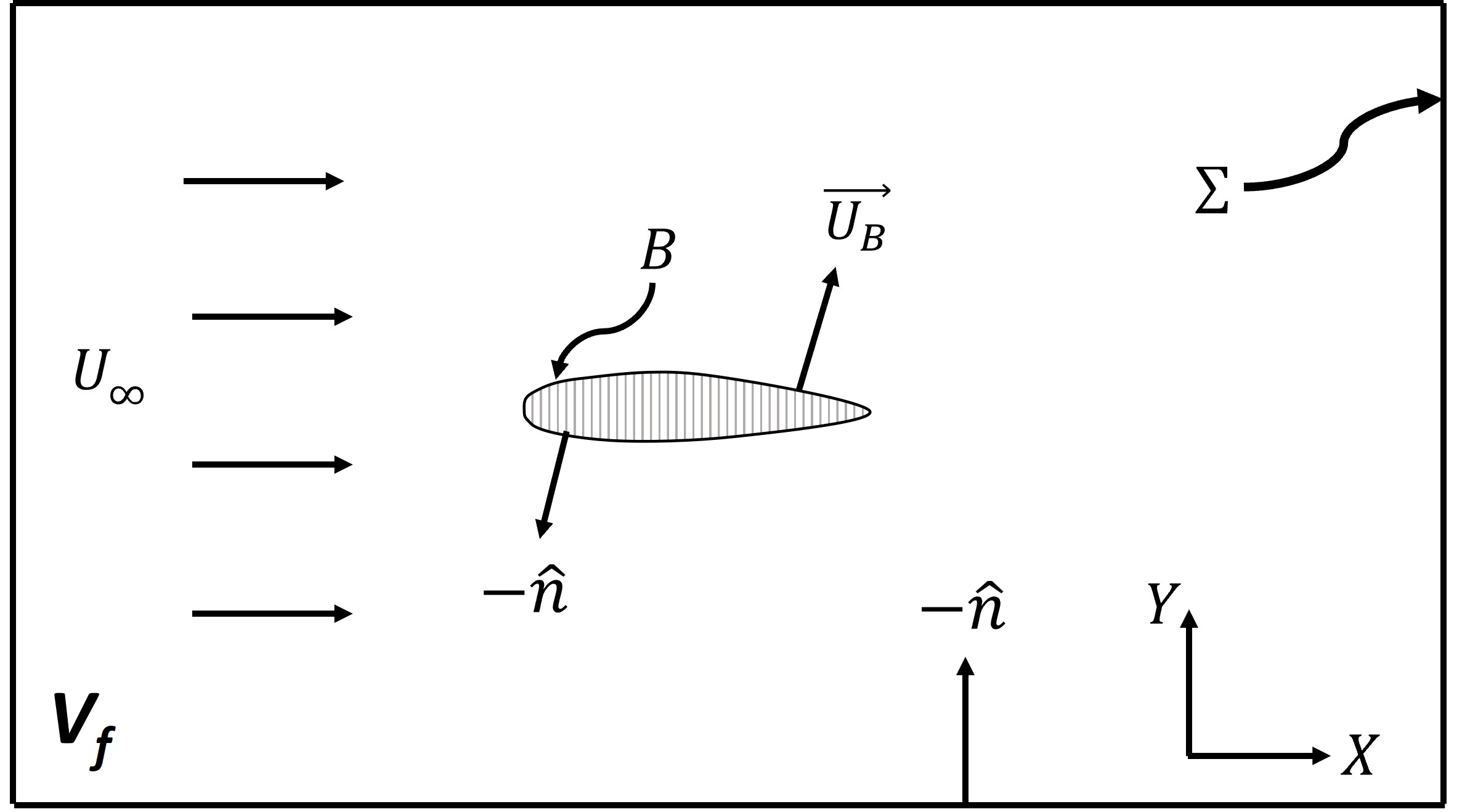}
    \caption{Schematic of the setup for the force-partitioning method (FPM), along with relevant symbols.}
    \label{fpm_schema}
\end{figure}

The Force Partitioning Method (FPM)~\cite{menon2021quantitative, menon2021significance} provides a systematic framework to decompose the pressure loads acting on an immersed body into distinct, physically interpretable components. Consider a body with surface $B$ immersed in a flow domain $V_f$, as illustrated in Fig. \ref{fpm_schema}. The fluid domain is externally bounded by the surface $\Sigma$. The pressure-induced force acting on the immersed surface in the $i^{\mathrm{th}}$ direction can then be expressed as a sum of individual contributions as follows:
\begin{align}
\begin{split}
    F_i =\int_B p n_i dS =  &\overbrace{-\rho \int_B \hat{n}.\Bigg(\frac{dU_B}{dt} \phi_i\Bigg)dS}^\text{\clap{$\text{F}_\text{AR}$}} \overbrace{-2 \rho \int_{V_f} Q\phi_i dV}^\text{\clap{$\text{F}_\text{VIF}$}} + \overbrace{\mu \int_{V_f} (\nabla^2 u).\nabla \phi_i dV}^\text{\clap{$\text{F}_\text{VIS}$}} \\
    &\overbrace{- \rho \int_{\Sigma} \hat{n}.\Bigg(\frac{du}{dt} \phi_i\Bigg) dS}^\text{\clap{$\text{F}_\text{BOUND}$}} \text{ for $i = 1,2$}    
\end{split}
\label{fpm_decomposition}
\end{align}
where $U_B$ denotes the local velocity of the immersed surface. The four terms on the right-hand side of Eq.~\ref{fpm_decomposition} represent distinct components of the pressure-induced force acting on the body’s surface. The first term, $\text{F}_\text{AR}$, corresponds to the acceleration-reaction force, which includes contributions from both the linear acceleration reaction~\cite{menon2021initiation} and the centripetal acceleration reaction~\cite{zhang2015centripetal}. The second term, $\text{F}_\text{VIF}$, represents the vortex-induced force, which arises from vortical and shear-layer regions in the flow. It involves the scalar quantity $Q$, defined as $Q = \tfrac{1}{2}\|\Omega\|^2 - \|S\|^2$, where $\Omega$ and $S$ denote the antisymmetric and symmetric parts of the velocity-gradient tensor ($\nabla \mathbf{u}$), respectively. This component often dominates in flows at relatively high Reynolds numbers~\cite{zhang2015centripetal, menon2021initiation, menon2021quantitative}. The third term, $\text{F}_\text{VIS}$, represents the pressure force associated with viscous diffusion within the flow, which becomes significant primarily at low Reynolds numbers. Finally, the fourth term, $\text{F}_\text{BOUND}$, accounts for the effect of flow acceleration at the outer boundary of the computational domain. This contribution is generally negligible for sufficiently large domains~\cite{zhang2015mechanisms} and in the presence of a steady freestream.

A key component of FPM is the ``influence field", denoted by $\phi_i$, which field a potential field obtained by solving the following Laplace equation 
\begin{align}
    \nabla^2\phi_i = 0, \quad \text{in $V_f$ with } \hat{n}.\nabla \phi_i =\Bigg\{
    \begin{split}
        & n_i, \text{on $B$} \\ 
        & 0,   \text{ on $\Sigma$}
    \end{split}
    \quad \text{for $i=1,2.$}
    \label{phi_equation}
\end{align}
for each force direction $i$ of interest. It is important to note that $\phi_i$ depends solely on the geometry of the immersed surface and is independent of the surrounding flow field. The potential $\phi_i$ can be computed for any given immersed body using standard numerical techniques. Additional details on the formulation and derivation of FPM  can be found in Refs. \citep{zhang2015centripetal,zhang2015mechanisms,menon2021significance,menon2021initiation,menon2021quantitative}. In the current problem, $i=1$ represent streamwise direction, which aligns with the direction of thrust on the wing and the influence field of interest is therefore $\phi_1$.

\section{\label{LEVBM_model} LEV Based Model }
\begin{figure}
    \centering
    \includegraphics[width=0.6\textwidth, trim={0.0cm 0cm 0cm 0cm}, clip]{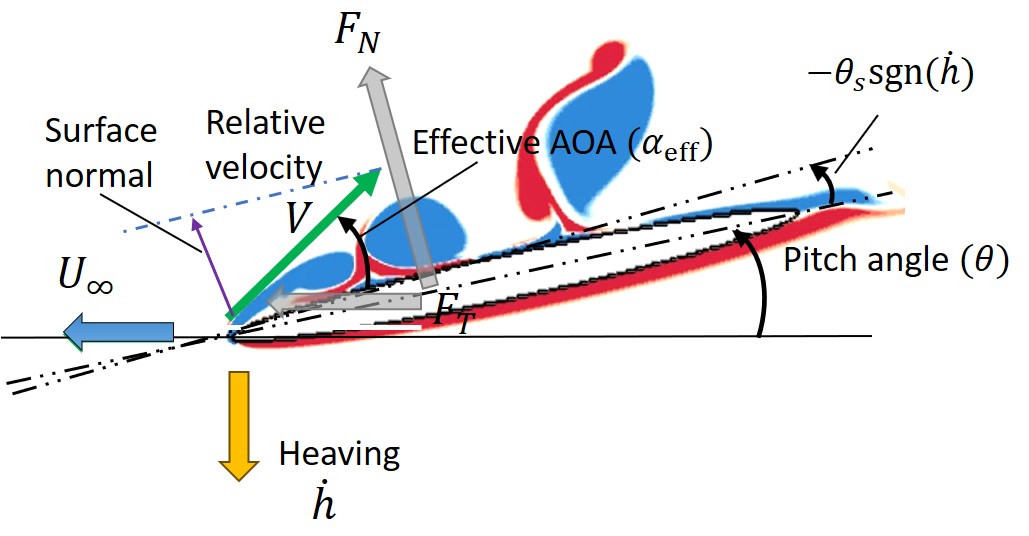}
    \caption{Schematic depicting the key features of the LEV based model.}
    \label{model_schema}
\end{figure}

In our previous work~\cite{raut2024hydrodynamic,seo2023hydrodynamics}, we demonstrated the dominant role of leading-edge vortices (LEVs) in thrust generation for wave-assisted propulsion (WAP) foils and proposed a new phenomenological model to describe LEV-induced thrust generation in flapping foils. The key aspects of this model are summarized here, as they provide the foundation for the design modifications explored in the present study. According to the Kutta–Joukowski theorem, the force acting normal to the foil, $F_N$ (see Fig.~\ref{model_schema}), can be expressed as:
\begin{equation}
    F_N = \rho V \Gamma
    \label{F_N}
\end{equation}
Here, $\Gamma$ denotes the circulation around the hydrofoil, and $V$ represents the instantaneous relative velocity of the foil with respect to the surrounding flow, defined as $V = \sqrt{U_\infty^2 + \dot{h}^2}$. Our previous analysis revealed that the circulation $\Gamma$ is primarily governed by the dynamics of the leading-edge vortex (LEV). The formation and strength of the LEV are proportional to the component of the incoming flow velocity at the leading edge that is normal to the foil surface. This component is determined by the effective angle of attack, $\alpha_\textrm{eff}$, which arises from both the oncoming flow and the heaving-pitching motion of the foil. Consequently, the circulation $\Gamma$ can be expressed as:
\begin{equation}
 \Gamma \propto  V \sin (\alpha_\textrm{eff})   
 \label{circulation}
\end{equation}
where the effective angle-of-attack from the surface of the foil, which is given by
\begin{equation}
  {\alpha _\textrm{eff}} (t) = {\tan ^{ - 1}}\left({V_\textrm{LE}(t)/U_\infty} \right)- \theta_\textrm{LE}(t)
  \label{alpha_eff}
\end{equation}
where the first term represents the angle of attack induced at the leading edge due to the combined heaving and pitching motions of the foil. The instantaneous lateral velocity of the leading edge is given by $V_\textrm{LE}(t) = -\left( \dot{h}(t) - \dot{\theta}(t) X_e \cos{\theta(t)} \right)$,
which accounts for the translational and rotational contributions to the motion.  
The second term, $\theta_\textrm{LE}$, denotes the local inclination of the foil surface at the leading edge, over which the LEV develops and imparts its force. It is expressed as $\theta_\textrm{LE}(t) = \theta(t) - \theta_s \, \mathrm{sgn}\left( \dot{h}(t) \right)$, where $\theta_s$ is a constant angular offset between the tangent to the surface near the leading edge and the foil chord line. This offset parameter $\theta_s$ depends on the local thickness and geometric shape of the foil near the leading edge.

Further assuming the following modified normalization of the normal force
\begin{equation}
    C_N = \frac{F_N}{\frac{1}{2}\rho V_\textrm{max}^2 C}
    \label{modified_thrust_coeff}
\end{equation}
where we employ the maximum value of $V$ instead of $U_\infty$ and using equation \ref{F_N}, \ref{circulation} and \ref{modified_thrust_coeff} we arrive at $C_N \propto \sin 
\left( \alpha_\textrm{eff} \right)$. From this it follows that the thrust coefficient satisfies the following proportionality
\begin{equation}
    C_T \propto \sin 
{\left( \alpha_\textrm{eff} \right)} \sin{\left( \theta_\textrm{LE} \right)}
    \label{eq_ctprop}
\end{equation}
Denoting the time-dependent function on the right-hand side, which depends solely on the kinematics and geometry of the foil, by $\Lambda_\textrm{LEV}$, the mean thrust predicted by this model can be expressed as:
\begin{equation}
    \bar{C}_T = K \bar{\Lambda}_\text{LEV} (\theta_s)
    \label{eq_ctbar}
\end{equation}
where $\Lambda_\text{LEV}(t;\theta_s)$ can be expanded (using Eq. \ref{alpha_eff}) as:
\begin{align}
    \Lambda_\text{LEV}(t;\theta_s) &=\sin 
{\left( \alpha_\textrm{eff} \right)} \sin{\left( \theta_\textrm{LE} \right)}=\sin {\left( {\tan ^{ - 1}}\left({V_\textrm{LE}(t)/U_\infty} \right)- \theta_\textrm{LE}(t) \right)} \sin{\left( \theta_\textrm{LE} \right)}\label{K_expanded}
\end{align}

For a given flapping foil, the above model for mean thrust has two unknown parameters: $K$, a proportionality constant, and $\theta_s$, which is related to the geometry of the foil near the leading edge. For the present thin elliptic foil, these parameters are determined to be $K = 5.0$ and $\theta_s = 0^\circ$.


Furthermore, as demonstrated in our previous work~\cite{raut2024hydrodynamic}, the above model predicts that the thrust is maximized for a pitching amplitude given by:
\begin{equation}
    \theta^\text{opt}_{0} = 0.5 \tan^{-1}\left( \pi \text{St}_w \right)-\theta_s
    \label{theta0max_eq}
\end{equation}
Thus, controlling the pitching amplitude is crucial for achieving maximum thrust, and the geometry of the leading edge, which influences $\theta_s$, also plays a significant role in determining the thrust performance.

\clearpage
\bibliography{references}

\begin{thebibliography}{35}
\expandafter\ifx\csname natexlab\endcsname\relax\def\natexlab#1{#1}\fi
\expandafter\ifx\csname bibnamefont\endcsname\relax
  \def\bibnamefont#1{#1}\fi
\expandafter\ifx\csname bibfnamefont\endcsname\relax
  \def\bibfnamefont#1{#1}\fi
\expandafter\ifx\csname citenamefont\endcsname\relax
  \def\citenamefont#1{#1}\fi
\expandafter\ifx\csname url\endcsname\relax
  \def\url#1{\texttt{#1}}\fi
\expandafter\ifx\csname urlprefix\endcsname\relax\def\urlprefix{URL }\fi
\providecommand{\bibinfo}[2]{#2}
\providecommand{\eprint}[2][]{\url{#2}}

\bibitem[{\citenamefont{Triantafyllou et~al.}(1993)\citenamefont{Triantafyllou, Triantafyllou, and Grosenbaugh}}]{TRIANTAFYLLOU1993205}
\bibinfo{author}{\bibfnamefont{G.}~\bibnamefont{Triantafyllou}}, \bibinfo{author}{\bibfnamefont{M.}~\bibnamefont{Triantafyllou}}, \bibnamefont{and} \bibinfo{author}{\bibfnamefont{M.}~\bibnamefont{Grosenbaugh}}, \bibinfo{journal}{Journal of Fluids and Structures} \textbf{\bibinfo{volume}{7}}, \bibinfo{pages}{205} (\bibinfo{year}{1993}), ISSN \bibinfo{issn}{0889-9746}, \urlprefix\url{https://www.sciencedirect.com/science/article/pii/S0889974683710121}.

\bibitem[{\citenamefont{Anderson et~al.}(1998)\citenamefont{Anderson, Streitlien, Barrett, and Triantafyllou}}]{anderson1998oscillating}
\bibinfo{author}{\bibfnamefont{J.~M.} \bibnamefont{Anderson}}, \bibinfo{author}{\bibfnamefont{K.}~\bibnamefont{Streitlien}}, \bibinfo{author}{\bibfnamefont{D.}~\bibnamefont{Barrett}}, \bibnamefont{and} \bibinfo{author}{\bibfnamefont{M.~S.} \bibnamefont{Triantafyllou}}, \bibinfo{journal}{Journal of Fluid mechanics} \textbf{\bibinfo{volume}{360}}, \bibinfo{pages}{41} (\bibinfo{year}{1998}).

\bibitem[{\citenamefont{Wu et~al.}(2020)\citenamefont{Wu, Zhang, Tian, Li, and Lu}}]{wu2020review}
\bibinfo{author}{\bibfnamefont{X.}~\bibnamefont{Wu}}, \bibinfo{author}{\bibfnamefont{X.}~\bibnamefont{Zhang}}, \bibinfo{author}{\bibfnamefont{X.}~\bibnamefont{Tian}}, \bibinfo{author}{\bibfnamefont{X.}~\bibnamefont{Li}}, \bibnamefont{and} \bibinfo{author}{\bibfnamefont{W.}~\bibnamefont{Lu}}, \bibinfo{journal}{Ocean Engineering} \textbf{\bibinfo{volume}{195}}, \bibinfo{pages}{106712} (\bibinfo{year}{2020}).

\bibitem[{\citenamefont{Srygley and Thomas}(2002)}]{srygley2002unconventional}
\bibinfo{author}{\bibfnamefont{R.}~\bibnamefont{Srygley}} \bibnamefont{and} \bibinfo{author}{\bibfnamefont{A.}~\bibnamefont{Thomas}}, \bibinfo{journal}{Nature} \textbf{\bibinfo{volume}{420}}, \bibinfo{pages}{660} (\bibinfo{year}{2002}).

\bibitem[{\citenamefont{Lauder and Drucker}(2002)}]{lauder2002forces}
\bibinfo{author}{\bibfnamefont{G.~V.} \bibnamefont{Lauder}} \bibnamefont{and} \bibinfo{author}{\bibfnamefont{E.~G.} \bibnamefont{Drucker}}, \bibinfo{journal}{Physiology} \textbf{\bibinfo{volume}{17}}, \bibinfo{pages}{235} (\bibinfo{year}{2002}).

\bibitem[{\citenamefont{Dickinson et~al.}(1999)\citenamefont{Dickinson, Lehmann, and Sane}}]{dickinson1999wing}
\bibinfo{author}{\bibfnamefont{M.~H.} \bibnamefont{Dickinson}}, \bibinfo{author}{\bibfnamefont{F.-O.} \bibnamefont{Lehmann}}, \bibnamefont{and} \bibinfo{author}{\bibfnamefont{S.~P.} \bibnamefont{Sane}}, \bibinfo{journal}{science} \textbf{\bibinfo{volume}{284}}, \bibinfo{pages}{1954} (\bibinfo{year}{1999}).

\bibitem[{\citenamefont{Xing and Yang}(2023)}]{xing2023wave}
\bibinfo{author}{\bibfnamefont{J.}~\bibnamefont{Xing}} \bibnamefont{and} \bibinfo{author}{\bibfnamefont{L.}~\bibnamefont{Yang}}, \bibinfo{journal}{Renewable and Sustainable Energy Reviews} \textbf{\bibinfo{volume}{184}}, \bibinfo{pages}{113589} (\bibinfo{year}{2023}).

\bibitem[{\citenamefont{Zhang et~al.}(2024{\natexlab{a}})\citenamefont{Zhang, Yue, Zhou, Gao, Zhang, and Chen}}]{zhang2024experimental}
\bibinfo{author}{\bibfnamefont{Y.}~\bibnamefont{Zhang}}, \bibinfo{author}{\bibfnamefont{J.}~\bibnamefont{Yue}}, \bibinfo{author}{\bibfnamefont{S.}~\bibnamefont{Zhou}}, \bibinfo{author}{\bibfnamefont{F.}~\bibnamefont{Gao}}, \bibinfo{author}{\bibfnamefont{W.}~\bibnamefont{Zhang}}, \bibnamefont{and} \bibinfo{author}{\bibfnamefont{W.}~\bibnamefont{Chen}}, \bibinfo{journal}{Applied Ocean Research} \textbf{\bibinfo{volume}{148}}, \bibinfo{pages}{104010} (\bibinfo{year}{2024}{\natexlab{a}}).

\bibitem[{\citenamefont{Zhang et~al.}(2022)\citenamefont{Zhang, Xu, Ding, and Hu}}]{zhang2022wave}
\bibinfo{author}{\bibfnamefont{Y.}~\bibnamefont{Zhang}}, \bibinfo{author}{\bibfnamefont{L.}~\bibnamefont{Xu}}, \bibinfo{author}{\bibfnamefont{Z.}~\bibnamefont{Ding}}, \bibnamefont{and} \bibinfo{author}{\bibfnamefont{M.}~\bibnamefont{Hu}}, \bibinfo{journal}{Ocean Engineering} \textbf{\bibinfo{volume}{266}}, \bibinfo{pages}{112802} (\bibinfo{year}{2022}).

\bibitem[{\citenamefont{Xu et~al.}(2024)\citenamefont{Xu, Jing, Liao, Tang, Ma, Liu, and Pang}}]{xu2024analysis}
\bibinfo{author}{\bibfnamefont{P.}~\bibnamefont{Xu}}, \bibinfo{author}{\bibfnamefont{B.}~\bibnamefont{Jing}}, \bibinfo{author}{\bibfnamefont{Y.}~\bibnamefont{Liao}}, \bibinfo{author}{\bibfnamefont{H.}~\bibnamefont{Tang}}, \bibinfo{author}{\bibfnamefont{T.}~\bibnamefont{Ma}}, \bibinfo{author}{\bibfnamefont{J.}~\bibnamefont{Liu}}, \bibnamefont{and} \bibinfo{author}{\bibfnamefont{S.}~\bibnamefont{Pang}}, \bibinfo{journal}{Applied Ocean Research} \textbf{\bibinfo{volume}{153}}, \bibinfo{pages}{104223} (\bibinfo{year}{2024}).

\bibitem[{\citenamefont{Qi et~al.}(2020)\citenamefont{Qi, Jiang, Jia, Zou, and Zhai}}]{qi2020effect}
\bibinfo{author}{\bibfnamefont{Z.}~\bibnamefont{Qi}}, \bibinfo{author}{\bibfnamefont{M.}~\bibnamefont{Jiang}}, \bibinfo{author}{\bibfnamefont{L.}~\bibnamefont{Jia}}, \bibinfo{author}{\bibfnamefont{B.}~\bibnamefont{Zou}}, \bibnamefont{and} \bibinfo{author}{\bibfnamefont{J.}~\bibnamefont{Zhai}}, \bibinfo{journal}{Journal of Marine Science and Engineering} \textbf{\bibinfo{volume}{8}}, \bibinfo{pages}{303} (\bibinfo{year}{2020}).

\bibitem[{\citenamefont{Zhang et~al.}(2024{\natexlab{b}})\citenamefont{Zhang, Han, Hu, Chen, Li, Gao, and Chen}}]{zhang2024dual}
\bibinfo{author}{\bibfnamefont{Y.}~\bibnamefont{Zhang}}, \bibinfo{author}{\bibfnamefont{X.}~\bibnamefont{Han}}, \bibinfo{author}{\bibfnamefont{Y.}~\bibnamefont{Hu}}, \bibinfo{author}{\bibfnamefont{X.}~\bibnamefont{Chen}}, \bibinfo{author}{\bibfnamefont{Z.}~\bibnamefont{Li}}, \bibinfo{author}{\bibfnamefont{F.}~\bibnamefont{Gao}}, \bibnamefont{and} \bibinfo{author}{\bibfnamefont{W.}~\bibnamefont{Chen}}, \bibinfo{journal}{Renewable Energy} \textbf{\bibinfo{volume}{222}}, \bibinfo{pages}{119956} (\bibinfo{year}{2024}{\natexlab{b}}).

\bibitem[{\citenamefont{Yang et~al.}(2019)\citenamefont{Yang, Shi, and Wang}}]{yang2019systematic}
\bibinfo{author}{\bibfnamefont{F.}~\bibnamefont{Yang}}, \bibinfo{author}{\bibfnamefont{W.}~\bibnamefont{Shi}}, \bibnamefont{and} \bibinfo{author}{\bibfnamefont{D.}~\bibnamefont{Wang}}, \bibinfo{journal}{Ocean Engineering} \textbf{\bibinfo{volume}{179}}, \bibinfo{pages}{361} (\bibinfo{year}{2019}).

\bibitem[{\citenamefont{Bretschneider}(1959)}]{bretschneider1959wave}
\bibinfo{author}{\bibfnamefont{C.~L.} \bibnamefont{Bretschneider}}, \emph{\bibinfo{title}{Wave variability and wave spectra for wind-generated gravity waves}}, \bibinfo{number}{118} (\bibinfo{publisher}{The Board}, \bibinfo{year}{1959}).

\bibitem[{\citenamefont{Hasselmann et~al.}(1973)\citenamefont{Hasselmann, Barnett, Bouws, Carlson, Cartwright, Enke, Ewing, Gienapp, Hasselmann, Kruseman et~al.}}]{hasselmann1973measurements}
\bibinfo{author}{\bibfnamefont{K.}~\bibnamefont{Hasselmann}}, \bibinfo{author}{\bibfnamefont{T.~P.} \bibnamefont{Barnett}}, \bibinfo{author}{\bibfnamefont{E.}~\bibnamefont{Bouws}}, \bibinfo{author}{\bibfnamefont{H.}~\bibnamefont{Carlson}}, \bibinfo{author}{\bibfnamefont{D.~E.} \bibnamefont{Cartwright}}, \bibinfo{author}{\bibfnamefont{K.}~\bibnamefont{Enke}}, \bibinfo{author}{\bibfnamefont{J.}~\bibnamefont{Ewing}}, \bibinfo{author}{\bibfnamefont{A.}~\bibnamefont{Gienapp}}, \bibinfo{author}{\bibfnamefont{D.}~\bibnamefont{Hasselmann}}, \bibinfo{author}{\bibfnamefont{P.}~\bibnamefont{Kruseman}}, \bibnamefont{et~al.}, \bibinfo{journal}{Ergaenzungsheft zur Deutschen Hydrographischen Zeitschrift, Reihe A}  (\bibinfo{year}{1973}).

\bibitem[{\citenamefont{Filippas and Belibassakis}(2014)}]{filippas2014hydrodynamic}
\bibinfo{author}{\bibfnamefont{E.}~\bibnamefont{Filippas}} \bibnamefont{and} \bibinfo{author}{\bibfnamefont{K.}~\bibnamefont{Belibassakis}}, in \emph{\bibinfo{booktitle}{International conference on offshore mechanics and arctic engineering}} (\bibinfo{organization}{American Society of Mechanical Engineers}, \bibinfo{year}{2014}), vol. \bibinfo{volume}{45509}, p. \bibinfo{pages}{V08AT06A023}.

\bibitem[{\citenamefont{Politis and Politis}(2014)}]{politis2014biomimetic}
\bibinfo{author}{\bibfnamefont{G.}~\bibnamefont{Politis}} \bibnamefont{and} \bibinfo{author}{\bibfnamefont{K.}~\bibnamefont{Politis}}, \bibinfo{journal}{Journal of fluids and structures} \textbf{\bibinfo{volume}{47}}, \bibinfo{pages}{139} (\bibinfo{year}{2014}).

\bibitem[{\citenamefont{Feng et~al.}(2025)\citenamefont{Feng, Wang, Jiang, Zhang, Deng, and Chang}}]{feng2025dynamic}
\bibinfo{author}{\bibfnamefont{Z.}~\bibnamefont{Feng}}, \bibinfo{author}{\bibfnamefont{H.}~\bibnamefont{Wang}}, \bibinfo{author}{\bibfnamefont{L.}~\bibnamefont{Jiang}}, \bibinfo{author}{\bibfnamefont{J.}~\bibnamefont{Zhang}}, \bibinfo{author}{\bibfnamefont{C.}~\bibnamefont{Deng}}, \bibnamefont{and} \bibinfo{author}{\bibfnamefont{Z.}~\bibnamefont{Chang}}, \bibinfo{journal}{Ocean Engineering} \textbf{\bibinfo{volume}{323}}, \bibinfo{pages}{120551} (\bibinfo{year}{2025}).

\bibitem[{\citenamefont{Raut et~al.}(2024)\citenamefont{Raut, Seo, and Mittal}}]{raut2024hydrodynamic}
\bibinfo{author}{\bibfnamefont{H.~S.} \bibnamefont{Raut}}, \bibinfo{author}{\bibfnamefont{J.-H.} \bibnamefont{Seo}}, \bibnamefont{and} \bibinfo{author}{\bibfnamefont{R.}~\bibnamefont{Mittal}}, \bibinfo{journal}{Journal of Fluid Mechanics} \textbf{\bibinfo{volume}{999}}, \bibinfo{pages}{A1} (\bibinfo{year}{2024}).

\bibitem[{\citenamefont{Raut et~al.}(2025{\natexlab{a}})\citenamefont{Raut, Seo, and Mittal}}]{raut2025dynamics}
\bibinfo{author}{\bibfnamefont{H.~S.} \bibnamefont{Raut}}, \bibinfo{author}{\bibfnamefont{J.-H.} \bibnamefont{Seo}}, \bibnamefont{and} \bibinfo{author}{\bibfnamefont{R.}~\bibnamefont{Mittal}}, \bibinfo{journal}{Ocean Engineering} \textbf{\bibinfo{volume}{317}}, \bibinfo{pages}{119930} (\bibinfo{year}{2025}{\natexlab{a}}).

\bibitem[{\citenamefont{Raut et~al.}(2025{\natexlab{b}})\citenamefont{Raut, Seo, and Mittal}}]{raut2025harnessing}
\bibinfo{author}{\bibfnamefont{H.~S.} \bibnamefont{Raut}}, \bibinfo{author}{\bibfnamefont{J.-H.} \bibnamefont{Seo}}, \bibnamefont{and} \bibinfo{author}{\bibfnamefont{R.}~\bibnamefont{Mittal}}, \bibinfo{journal}{Phys. Rev. Fluids} \textbf{\bibinfo{volume}{10}}, \bibinfo{pages}{084705} (\bibinfo{year}{2025}{\natexlab{b}}), \urlprefix\url{https://link.aps.org/doi/10.1103/w9zp-q3gj}.

\bibitem[{\citenamefont{Wang et~al.}(2019)\citenamefont{Wang, Tian, Lu, Hu, and Luo}}]{wang2019dynamic}
\bibinfo{author}{\bibfnamefont{P.}~\bibnamefont{Wang}}, \bibinfo{author}{\bibfnamefont{X.}~\bibnamefont{Tian}}, \bibinfo{author}{\bibfnamefont{W.}~\bibnamefont{Lu}}, \bibinfo{author}{\bibfnamefont{Z.}~\bibnamefont{Hu}}, \bibnamefont{and} \bibinfo{author}{\bibfnamefont{Y.}~\bibnamefont{Luo}}, \bibinfo{journal}{Applied Mathematical Modelling} \textbf{\bibinfo{volume}{66}}, \bibinfo{pages}{77} (\bibinfo{year}{2019}).

\bibitem[{\citenamefont{Yang et~al.}(2018)\citenamefont{Yang, Shi, Zhou, Guo, and Wang}}]{yang2018numerical}
\bibinfo{author}{\bibfnamefont{F.}~\bibnamefont{Yang}}, \bibinfo{author}{\bibfnamefont{W.}~\bibnamefont{Shi}}, \bibinfo{author}{\bibfnamefont{X.}~\bibnamefont{Zhou}}, \bibinfo{author}{\bibfnamefont{B.}~\bibnamefont{Guo}}, \bibnamefont{and} \bibinfo{author}{\bibfnamefont{D.}~\bibnamefont{Wang}}, \bibinfo{journal}{Ocean Engineering} \textbf{\bibinfo{volume}{164}}, \bibinfo{pages}{127} (\bibinfo{year}{2018}).

\bibitem[{\citenamefont{Mittal et~al.}(2008)\citenamefont{Mittal, Dong, Bozkurttas, Najjar, Vargas, and Von~Loebbecke}}]{mittal2008versatile}
\bibinfo{author}{\bibfnamefont{R.}~\bibnamefont{Mittal}}, \bibinfo{author}{\bibfnamefont{H.}~\bibnamefont{Dong}}, \bibinfo{author}{\bibfnamefont{M.}~\bibnamefont{Bozkurttas}}, \bibinfo{author}{\bibfnamefont{F.}~\bibnamefont{Najjar}}, \bibinfo{author}{\bibfnamefont{A.}~\bibnamefont{Vargas}}, \bibnamefont{and} \bibinfo{author}{\bibfnamefont{A.}~\bibnamefont{Von~Loebbecke}}, \bibinfo{journal}{Journal of Computational Physics} \textbf{\bibinfo{volume}{227}}, \bibinfo{pages}{4825} (\bibinfo{year}{2008}).

\bibitem[{\citenamefont{Seo and Mittal}(2011)}]{seo2011sharp}
\bibinfo{author}{\bibfnamefont{J.~H.} \bibnamefont{Seo}} \bibnamefont{and} \bibinfo{author}{\bibfnamefont{R.}~\bibnamefont{Mittal}}, \bibinfo{journal}{Journal of computational physics} \textbf{\bibinfo{volume}{230}}, \bibinfo{pages}{7347} (\bibinfo{year}{2011}).

\bibitem[{\citenamefont{Seo et~al.}(2023)\citenamefont{Seo, Raut, and Mittal}}]{seo2023hydrodynamics}
\bibinfo{author}{\bibfnamefont{J.-H.} \bibnamefont{Seo}}, \bibinfo{author}{\bibfnamefont{H.}~\bibnamefont{Raut}}, \bibnamefont{and} \bibinfo{author}{\bibfnamefont{R.}~\bibnamefont{Mittal}}, \bibinfo{journal}{Bulletin of the American Physical Society}  (\bibinfo{year}{2023}).

\bibitem[{\citenamefont{Prakhar et~al.}(2025)\citenamefont{Prakhar, Seo, and Mittal}}]{prakhar2025bioinspired}
\bibinfo{author}{\bibfnamefont{S.}~\bibnamefont{Prakhar}}, \bibinfo{author}{\bibfnamefont{J.-H.} \bibnamefont{Seo}}, \bibnamefont{and} \bibinfo{author}{\bibfnamefont{R.}~\bibnamefont{Mittal}}, \bibinfo{journal}{Bioinspiration \& Biomimetics} \textbf{\bibinfo{volume}{20}}, \bibinfo{pages}{046009} (\bibinfo{year}{2025}).

\bibitem[{\citenamefont{Kumar et~al.}(2025)\citenamefont{Kumar, Seo, and Mittal}}]{kumar2025computational}
\bibinfo{author}{\bibfnamefont{S.}~\bibnamefont{Kumar}}, \bibinfo{author}{\bibfnamefont{J.-H.} \bibnamefont{Seo}}, \bibnamefont{and} \bibinfo{author}{\bibfnamefont{R.}~\bibnamefont{Mittal}}, \bibinfo{journal}{Journal of Fluid Mechanics} \textbf{\bibinfo{volume}{1010}}, \bibinfo{pages}{A53} (\bibinfo{year}{2025}).

\bibitem[{\citenamefont{Kraus}(2012)}]{kraus2012wave}
\bibinfo{author}{\bibfnamefont{N.~D.} \bibnamefont{Kraus}}, Ph.D. thesis, \bibinfo{school}{[Honolulu]:[University of Hawaii at Manoa],[May 2012]} (\bibinfo{year}{2012}).

\bibitem[{\citenamefont{Zhang et~al.}(2015)\citenamefont{Zhang, Hedrick, and Mittal}}]{zhang2015centripetal}
\bibinfo{author}{\bibfnamefont{C.}~\bibnamefont{Zhang}}, \bibinfo{author}{\bibfnamefont{T.~L.} \bibnamefont{Hedrick}}, \bibnamefont{and} \bibinfo{author}{\bibfnamefont{R.}~\bibnamefont{Mittal}}, \bibinfo{journal}{PloS one} \textbf{\bibinfo{volume}{10}}, \bibinfo{pages}{e0132093} (\bibinfo{year}{2015}).

\bibitem[{\citenamefont{Menon and Mittal}(2021{\natexlab{a}})}]{menon2021quantitative}
\bibinfo{author}{\bibfnamefont{K.}~\bibnamefont{Menon}} \bibnamefont{and} \bibinfo{author}{\bibfnamefont{R.}~\bibnamefont{Mittal}}, \bibinfo{journal}{Journal of Computational Physics} \textbf{\bibinfo{volume}{443}}, \bibinfo{pages}{110515} (\bibinfo{year}{2021}{\natexlab{a}}).

\bibitem[{\citenamefont{Zhou et~al.}(2025)\citenamefont{Zhou, Seo, and Mittal}}]{zhou2025hydrodynamically}
\bibinfo{author}{\bibfnamefont{J.}~\bibnamefont{Zhou}}, \bibinfo{author}{\bibfnamefont{J.-H.} \bibnamefont{Seo}}, \bibnamefont{and} \bibinfo{author}{\bibfnamefont{R.}~\bibnamefont{Mittal}}, \bibinfo{journal}{Journal of Fluid Mechanics} \textbf{\bibinfo{volume}{1014}}, \bibinfo{pages}{A32} (\bibinfo{year}{2025}).

\bibitem[{\citenamefont{Menon and Mittal}(2021{\natexlab{b}})}]{menon2021significance}
\bibinfo{author}{\bibfnamefont{K.}~\bibnamefont{Menon}} \bibnamefont{and} \bibinfo{author}{\bibfnamefont{R.}~\bibnamefont{Mittal}}, \bibinfo{journal}{Journal of Fluid Mechanics} \textbf{\bibinfo{volume}{918}}, \bibinfo{pages}{R3} (\bibinfo{year}{2021}{\natexlab{b}}).

\bibitem[{\citenamefont{Menon and Mittal}(2021{\natexlab{c}})}]{menon2021initiation}
\bibinfo{author}{\bibfnamefont{K.}~\bibnamefont{Menon}} \bibnamefont{and} \bibinfo{author}{\bibfnamefont{R.}~\bibnamefont{Mittal}}, \bibinfo{journal}{Journal of Fluid Mechanics} \textbf{\bibinfo{volume}{907}}, \bibinfo{pages}{A37} (\bibinfo{year}{2021}{\natexlab{c}}).

\bibitem[{\citenamefont{Zhang}(2015)}]{zhang2015mechanisms}
\bibinfo{author}{\bibfnamefont{C.}~\bibnamefont{Zhang}}, Ph.D. thesis, \bibinfo{school}{Johns Hopkins University} (\bibinfo{year}{2015}).

\end{thebibliography}
\end{document}